\documentclass[aps,prc,twocolumn,nofootinbib,superscriptaddress,showpacs,preprintnumbers,notitlepage]{revtex4-1}
\usepackage{amsmath,amsfonts,amssymb}
\usepackage[english]{babel}
\usepackage{hyperref}
\usepackage{comment}
\usepackage{soul}
\usepackage{color}
\usepackage{graphicx}
\usepackage{stackengine}
\usepackage{soul}
\usepackage{tikz}
\newcommand{\tcdr}[1]{\textcolor{black}{#1}}
\usepackage[markup=underlined]{changes}
\definechangesauthor[color=blue]{JR}
\definechangesauthor[color=green]{WC}

\begin{document}

\title{Isospin composition of the high-momentum fluctuations in nuclei from asymptotic momentum distributions}
\author{Jan Ryckebusch}%
\affiliation{Department of Physics and Astronomy, Ghent University, B-9000 Ghent, Belgium}
%
\author{Wim Cosyn}
\affiliation{Department of Physics and Astronomy, Ghent University, B-9000 Ghent, Belgium}
\author{Tom Vieijra}
\affiliation{Department of Physics and Astronomy, Ghent University, B-9000 Ghent, Belgium}
\author{Corneel Casert}
\affiliation{Department of Physics and Astronomy, Ghent University, B-9000 Ghent, Belgium}
%


\date{\today}

\begin{abstract}
\begin{description}
\item[Background]
High-momentum nucleons in a nuclear environment can be associated with short-range correlations (SRC) that primarily occur  between nucleon pairs. 
Observations and theoretical developments have indicated that the   
SRC properties can be captured by general quantitative principles that are subject to model-dependence upon quantification.
The variations in the aggregated effect of SRC across nuclei, however, can be quantified in an approximately model-independent fashion in terms of the so-called SRC scaling factors that capture the aggregated effect of SRC for a specific nucleus $A$ relative to the deuteron ($A$-to-$d$). 
\item[Purpose] We aim to provide predictions for the SRC scaling factors across the nuclear periodic table and determine the relative contribution of the different nucleon pair combinations to this quantity. We will also determine the SRC scaling factors for both bound protons and bound neutrons and study how these quantities evolve with the neutron-to-proton ($\frac{N}{Z}$) ratio in asymmetric nuclei.
\item[Methods] We employ the low-order correlation operator approximation (LCA) to compute the SRC contribution to the single-nucleon momentum distribution and ratios of $A$-to-$d$ momentum distributions. We do this for a sample of fifteen nuclei from He  to Pb thereby gaining access to the evolution of the SRC scaling factor with the nuclear mass $4 \le A \le 208$ and the neutron-to-proton ratio $1.0 \le \frac{N}{Z} \le 1.54$. 
\item[Results] We provide evidence for approximate $A$-to-$d$ scaling of the single-nucleon momentum distribution at nucleon momenta exceeding about 4~fm$^{-1}$.  For the studied sample of fifteen nuclei, the total SRC scaling factor is in the range 4.05-5.14 of which roughly 3 can be attributed to proton-neutron (pn) correlations. The SRC scaling factors receive sizable contributions from  pp and nn correlations.  They depend on the $\left( \frac{N}{Z} \right)$ ratio reflecting the fact that the minority species (protons) becomes increasingly more short-range correlated with increasing $\left( \frac{N}{Z} \right)$.  We compare the computed SRC scaling factors in the LCA with those of ab-initio calculations and with measured quantities from SRC-sensitive inclusive electron-scattering data.  
\item[Conclusions] It is shown that the LCA provides predictions for the SRC scaling factors across the nuclear table that are in line with measured values. In asymmetric nuclei there are sizable differences between the SRC scaling factors for protons and neutrons. It is suggested that this phenomenon may impact the variations of the magnitude of the European Muon Collaboration (EMC) effect across nuclei. Our results corroborate the finding that SRC physics can be qualitatively understood by universal principles that build on local modifications of mean-field wave functions of nucleon pairs. 
\end{description}
\end{abstract}

\maketitle
%
\section{Introduction}
\label{sec:intro}
Nuclear short-range correlations (SRC) are a primary source of high-momentum and high-energy spatio-temporal fluctuations in atomic nuclei. They are connected to nucleon-pair correlations in nuclei and induce dynamical effects that go beyond the independent-nucleon picture of atomic nuclei~\cite{Frankfurt:2008zv, Arrington:2011xs, Atti:2015eda, Hen:2016kwk}. Throughout the last decade an improved quantitative understanding of SRC has been accomplished thanks to concerted experimental efforts in  
exclusive and semi-exclusive electron-scattering reactions of nuclei under peculiar kinematics. The analysis of  $A(e,e^{\prime}pp)$,  $A(e,e'np)$ \cite{Duer:2018sxh} and $A(e,e^{\prime}N)$ reactions for example has provided detailed information on the isospin dependence~\cite{Piasetzky:2006ai,Subedi:2008zz,Hen:2014nza,Duer2018,Duer:2018sxh}, on the quantum numbers~\cite{Colle:2015ena}, and on the center-of-mass motion~\cite{Colle:2013nna,Cohen:2018gzh}  of short-range correlated nucleon pairs.    

There are also two known classes of inclusive electron-scattering $A(e,e^{\prime})$ reactions that have been connected to SRC.  In both situations the aggregated impact of SRC in nucleus $A$ is determined relative to the deuteron $d$ and involves the observation of a scaling mechanism of the ratio of the cross sections on A relative to $d$. Both classes, however, refer to different resolution scales and nucleon-momentum conditions. These physical conditions are commonly quantified by the virtuality $Q^2 = -q_{\mu} q ^{\mu}$ of the exchanged virtual photon (four-momentum $q^{\mu}(\omega, \vec{q})$ in the laboratory frame) in the electron-nucleus interaction, and the Lorentz scalar known as the Bjorken-Feynman variable $x = \frac{Q^2}{2M_N \omega}$, with $M_N$ the nucleon mass. 
\begin{itemize}
\item First, it has been observed that in well-selected kinematics~\cite{Frankfurt:1993sp}---namely sufficiently small resolution scales and  $x$ values $1.5 \lesssim x \lesssim 1.9$ that single out virtual-photon absorption on nucleon pairs--- the $A$-to-$d$ $(e,e^{\prime})$ cross sections approximately scale. 
The extraction of the scaling factor 
\begin{equation}
a_2^{exp}(A) = \frac {2}{A} \frac
{\sigma^{A}(e,e^{\prime})}
{\sigma^{d}(e,e^{\prime})}   \; \; \; 
\left( 1.5 \lesssim x \lesssim 1.9 \; ; \; Q^2 \approx 2~\text{GeV}^2 \right) \; ,
\label{eq:a2meas}
\end{equation}
has been the subject of intense experimental campaigns. To our knowledge, the $a_2^{exp}(A)$ have been measured \cite{PhysRevLett.96.082501,Fomin:2011ng,Schmookler:2019nvf} for  nine target nuclei:  $ ^{3}$He, $^{4}$He, $^{9}$Be, $^{12}$C, $^{27}$Al, $^{56}$Fe, $^{63}$Cu, $^{197}$Au, $^{208}$Pb.
\item The European Muon Collaboration (EMC) effect refers to the observation that at resolution scales that probe partons ($Q^2 \gtrsim 5$~GeV$^2$ ) and conditions $0.2 \lesssim x \lesssim 0.7$ (moderate- to high-momentum quarks), the ratio of the nucleon-weighted cross sections $\frac {2}{A} \frac
{\sigma^{A}(e,e^{\prime})}
{\sigma^{d}(e,e^{\prime})} $ depends on the target nucleus. This observation is commonly parameterized by means of the quantity
\begin{eqnarray}\label{eq:sizeEMC}
b_2^{exp} (A) & \equiv & - \frac {d R_{EMC} (A,x)} {d x} \nonumber \\
& = &
- \frac {d \left( \frac{2F_2 ^{A} (x, Q^2)}{ A F_2 ^{d} (x, Q^2)} \right)} {d x}
\; \; \; 
\nonumber \\
& & \left( 0.2 \lesssim x \lesssim 0.7 \; ; \; Q^2  \gtrsim 5~ \text{GeV}^2 \right) \; ,
\end{eqnarray}
where one has that $\sigma^{A}(e,e^{\prime}) \sim F_2 ^{A} (x, Q^2)$. The $b_2^{exp}(A)$ have been measured for the same nine target nuclei for which $a_2^{exp}(A)$ data are available. It came as a rather big surprise \cite{Hen:2012fm,Weinstein:2010rt,Arrington:2012ax} that within the error bars the size of the EMC effect parameterized by $b_2^{exp}(A)$ is roughly linearly correlated with the measured values of $a_2^{exp}(A)$.     
\end{itemize}

%

As it is inherently challenging to compute the coefficients $b_2^{exp}(A)$ and  $a_2^{exp}(A)$ from ratios of computed $\sigma^{A}(e,e^{\prime})$ and $\sigma^{d}(e,e^{\prime})$ cross sections in selected but large ranges of phase space, 
one has resorted to alternate techniques to gain theoretical access to their values. It has been argued that theoretical predictions for the $a_2^{exp}(A)$ (and indirectly for the $b_2^{exp}(A)$) can be obtained by evaluating ratios of bound-nucleon probability distributions  in the limits of vanishing relative distance $r_{12}$, or equivalently, infinitely high relative momentum $p_{12}$
\begin{eqnarray}
a_2 (A) & = &     \lim _{r_{12} \to 0} \frac {\rho ^ A (r_{12} , \Lambda) }  {\rho ^ d (r_{12},  \Lambda)} \; ,
\label{eq:a2fromasymptotics_r}
\\
a_2 (A) & = &     \lim _{p_{12} \to \infty } \frac {n ^ A (p_{12}, \Lambda) }  {n ^ d(p_{12}, \Lambda)} \; .
\label{eq:a2fromasymptotics_p}
\end{eqnarray}
Here, the $\rho^{A}(r_{12}, \Lambda)r_{12}^2dr_{12}$ is related to the probability of finding a nucleon pair in A with a relative separation $r_{12} = \left| \vec{r}_1 - \vec{r}_2 \right|$ in the interval $[r_{12}, r_{12} + d r_{12}]$. Similarly, the $n^{A}(p_{12},\Lambda)p_{12}^2dp_{12}$ is related to the probability of finding a nucleon pair in A with a relative momentum $p_{12} = \left| \vec{p}_1 - \vec{p}_2 \right|$ in the interval $[p_{12}, p_{12} + d p_{12}]$.  The validity of the Eqs.~(\ref{eq:a2fromasymptotics_r}) and (\ref{eq:a2fromasymptotics_p}) is very much based on the idea that the very short internucleon behavior in nuclei is characterized by universal functions that simply differ across nuclei by a scaling factor that relates to the measured $a_2^{exp}(A)$. In coordinate space this property can be captured by the factorization expression
\begin{eqnarray} \label{eq:shortdisfuncs}
\phantom{aa}    & \phantom{=} & \rho ^ A _{NN^{\prime} \in \{\text{pn, pp, nn}\} } (r_{12} \lesssim r_{\Lambda}, \Lambda)  \nonumber \\ 
    &\approx& C^A_{NN^{\prime}} (\Lambda)  \left| \psi_{NN^{\prime}} (r_{12}, \Lambda)  \right|^2 \;.
\end{eqnarray}
\tcdr{The $r_{\Lambda}$ is of the order of 1~{fm}. Its precise value is connected with the ultraviolet regulator scale $\Lambda$ implicit for a particular nucleon-nucleon interaction model and the larger $\Lambda$ is the smaller $r_{\Lambda}$ is~\cite{Lynn:2019vwp}.}  Further, the variation across nuclei is contained in the factor $C^A_{NN^{\prime}} (\Lambda)$, whereby the index $NN^{\prime}$ accounts for variations in the scaling factors across the different types of nucleon pairs. The quantities $C^A_{NN^{\prime}} (\Lambda)$ are often referred to as the ``contacts'' for $NN^{\prime}$ pairs~\cite{Alvioli:2016wwp,Weiss:2015mba,Weiss:2016obx,Cruz-Torres:2019fum}.  The distributions in  Eqs.~(\ref{eq:a2fromasymptotics_r}) and (\ref{eq:a2fromasymptotics_p}) are model-dependent \cite{Bogner:2012zm, Lynn:2019vwp}, an aspect that is highlighted by the label $\Lambda$. By evaluating $A$-to-$d$ ratios as in Eqs.~(\ref{eq:a2fromasymptotics_r}) and (\ref{eq:a2fromasymptotics_p}), however, one can gain access to quantities that are approximately model-independent and forge connections with measured quantities. \tcdr{The model independence of the ratio of Eqs.~(\ref{eq:a2fromasymptotics_r}) can be intuitively understood by realizing that for a given nucleon-nucleon interaction the highly local positional neighborhood of a nucleon in nucleus $A$ is not very different from the one of a nucleon in the deuteron.}

A major challenge is to isolate the generative mechanisms in the scaling factors $a_2^{exp}(A)$ and $b_2^{exp}(A)$. For example, the isospin dependence of the size of the EMC effect $b_2^{exp}(A)$ provides access to the important issue of the flavor dependence in nuclear quark distributions~\cite{Cloet:2009qs, Cloet:2015tha}, and has been a subject of recent debates~\cite{Schmookler:2019nvf, Arrington:2019wky,Hen:2019jzn}.  Access to these issues can be gained from determining the contribution of the different nucleon pair combinations to the short-distance modifications of nucleons embedded in a nuclear environment. This is the major topic of investigation in this paper.  


Ab-initio low-energy nuclear theory has been used to compute the  $a_2 (A)$ for a number of light nuclei with $A \le 40$~\cite{Chen:2016bde, Lynn:2019vwp}.  The calculations use advanced importance-sampling  based quantum many-body theories~\cite{Carlson:2014vla, Wiringa:2013ala, Lonardoni:2017egu} in combination with various forms of nucleon-nucleon ($NN$) interactions to determine the $ \lim _{r_{12} \to 0} \frac {\rho ^ A (r_{12}, \Lambda) }  {\rho ^ d (r_{12}, \Lambda)}$ of Eq.~(\ref{eq:a2fromasymptotics_r}). These calculations have also shed light on the appropriateness of the expression (\ref{eq:a2fromasymptotics_r}) and the sensitivity to the adopted model ``$\Lambda$''. The $^{40}$Ca result in Figure~3 of \cite{Lynn:2019vwp} has illustrated that the proposed $A$-to-$d$ scaling for the relative density distribution at very short internucleon distances is approximate. Also the  shrinking size of phase space for $ \lim _{r_{12} \to 0}$ poses challenges for the importance-sampling techniques.     
The (generalized) contact formalism~\cite{Weiss:2015mba,Weiss:2016obx,Cruz-Torres:2017sjy, Alvioli:2016wwp,Cruz-Torres:2019fum} builds on the Eq.~(\ref{eq:shortdisfuncs}) to construct pair-density functions and correlation functions for the pp, nn and pn pairs.  The contact formalism  can  be applied to  heavy nuclei ($A>40$) but requires input  either from data \cite{Duer:2018sxh} or from computed momentum distributions~\cite{Weiss:2015mba,Cruz-Torres:2019fum}. As $A>40$ ab-initio calculations are not available, no systematic predictions for the SRC scaling factor for medium-heavy and heavy nuclei have been produced so far. 

\begin{figure*}[ht!]
    \centering
    \includegraphics[width=\textwidth]{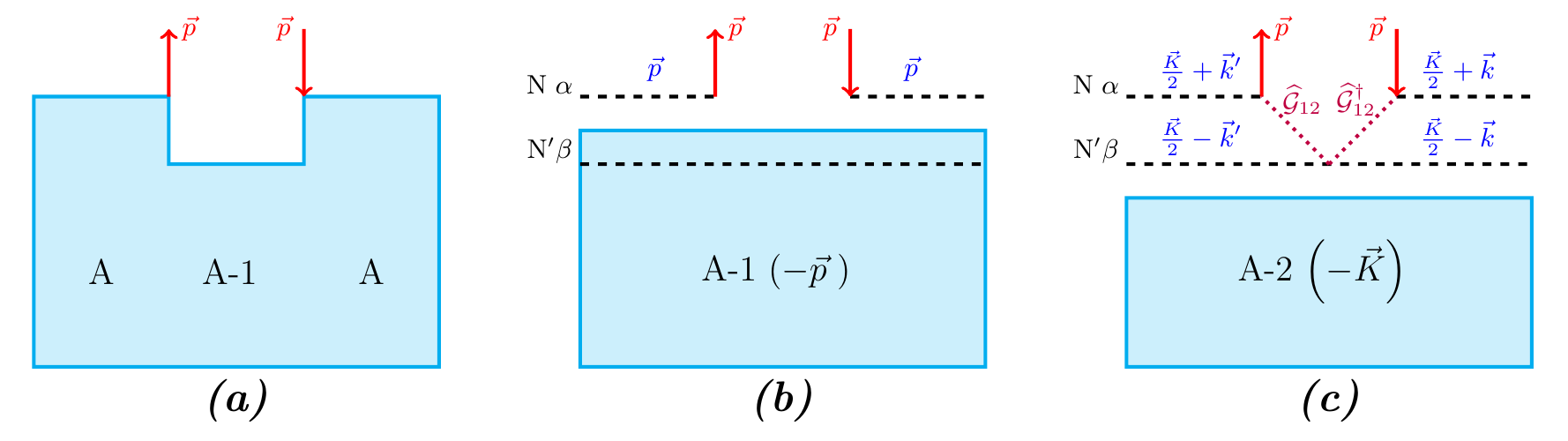}
    \caption{Schematic representation of the dominant contributions to the single-nucleon momentum distribution $n^{A}(\vec{p})$ (defined in diagram (a)) in LCA.
    \tcdr{The $n^{A}(\vec{p})$  quantifies the probability of removing
from the nuclear ground state a momentum $\vec{p}$ at a certain location $\vec{r}$ and putting it instantly back at another location $\vec{r}^{\; \prime}$ for all possible combinations of $\vec{r}$ and $\vec{r}^{\; \prime}$. }
    The black dashed lines denote IPM nucleons: \tcdr{they are characterized by their isospin $N, N^{\prime} \in \left\{ p, n \right\}$ and other IPM quantum numbers $\alpha, \beta$}. The purple dotted  lines denote the correlation operators in Eq.~(\ref{eq:SRCoperator}). Diagram (b) is the IPM contribution \tcdr{expressed in the format of  Eq.~(\ref{eq:IPMcontri_re}). The IPM contribution} dominates for $p<p_F$. Diagram (c) represents one of the SRC contributions between nucleon pairs \tcdr{(see Eq.~(\ref{eq:SRCcontri}))} and provides the bulk of the strength to $n^{A}(\vec{p})$ for $p>p_F$.}
    \label{fig:schememomendis}
\end{figure*}

The low-order correlation operator approximation (LCA) as proposed in \cite{Vanhalst:2014cqa, Ryckebusch:2018rct} is an alternate approximate method to compute the impact of SRC on nuclear momentum distributions. Recently, we have shown~\cite{Ryckebusch:2018rct} that LCA can reproduce the major trends of the observed $N/Z$ dependence of SRC~\cite{Duer2018}.  In line with the results of alternate calculations~\cite{PhysRevC.79.064308, Sargsian:2012sm}, LCA accounts for the fact that through the operation of the tensor force, the minority component (protons) is  substantially  more correlated than the majority component (neutrons) in asymmetric nuclei. 

In this work we present a systematic study of the SRC scaling factors based on the asymptotic high-momentum behavior of single-nucleon momentum distributions computed in LCA. We include fifteen nuclei in our study, including eight for which $a_2^{exp} (A)$ and $b_2^{exp} (A)$ data are available.  One of the major goals of the presented study is to uncover the trends in the isospin (flavor) composition of the high-momentum (short-distance) behavior of nuclei. To this end, we have included both symmetric ($\frac{N}{Z}=1$) and asymmetric ($\frac{N}{Z}>1$) nuclei providing a window on asymmetric neutron-rich matter~\cite{li2018nucleon}. The selection of nuclei was not random but was made on the basis of reaching a good coverage of both the  mass dependence $(4 \le A \le 208)$ and the neutron-to-proton dependence $(1 \le \frac{N}{Z} \le 1.54)$ of SRC. The four symmetric nuclei that are contained in our study are: $^{4}$He, $^{12}$C, $^{16}$O, $^{40}$Ca. The eleven asymmetric ones are  $^{9}$Be,  $^{27}$Al, $^{40}$Ar, $^{48}$Ca, $^{56}$Fe, $^{63}$Cu, $^{84}$Kr, $^{108}$Ag, $^{124}$Xe, $^{197}$Au, $^{208}$Pb.  With the presented calculations we can also address questions like (i) the degree of validity of the scaling behavior of the Eq.~(\ref{eq:a2fromasymptotics_p}) and, (ii) to what extent do the short-distance modifications affect protons and neutrons differently in asymmetric nuclear matter.   


In what follows, Sec.~\ref{subsec:LCAsinglenucleon} introduces the LCA method for computing the SRC contribution to single-nucleon momentum distributions. In Section~\ref{subsec:tailpartofobmd} we discuss the pair composition of the SRC part of the nucleon momentum distributions and forge connections to measured quantities from exclusive electroinduced two-nucleon knockout.  In Sec.~\ref{subsec:asymptotics_and_SRCscaling} we proceed with presenting and discussing the LCA results for the SRC scaling factors for 15 nuclei. We have included checks and balances and compared the computed SRC scaling factors with both data and theoretical results of ab-initio calculations.  We also conduct robustness checks \tcdr{of the presented methodology by testing the sensitivity of the SRC scaling factors to the momentum range that determines the ``asymptotic part'' of the single-nucleon momentum distribution.}   
Section~\ref{subsec:SRCscalingfactor} focuses on the differences in the SRC scaling factors for proton and neutrons in asymmetric nuclei. In Sec.~\ref{subsec:sizeEMCeffect} we exploit the conjectured relationship between the size of the EMC effect and the SRC scaling factors to shed light on the  isospin dependence of the underlying (unknown) generative mechanisms.

%
%
\section{Formalism and results}
\label{sec:formalism}
\subsection{Single-nucleon momentum distributions}
\label{subsec:LCAsinglenucleon}

The LCA is a methodology with applications in nuclear reactions and nuclear structure. In LCA one can compute the observables for SRC dominated nucleon knockout reactions ~\cite{Colle:2015ena, Ryckebusch:1997gn,  VanCuyck:2016fab, Colle:2015lyl, Stevens:2017orj}. Furthermore, the impact of SRC on nuclear momentum distributions \cite{Vanhalst:2014cqa, Ryckebusch:2018rct} can be quantified across the nuclear mass range because even for the heaviest nuclei the numerical calculations are manageable.  Central to the results of this work is the single-nucleon momentum distribution  that is generally defined as
\begin{equation}
n^{A}(\vec{p}) \sim \left< \Psi _A \right| a_{\vec{p}}^{\dagger} \; a_{\vec{p}}^{\phantom{\dagger}} \left| \Psi_A \right> \; ,
\end{equation}
with $\left| \Psi_A \right>$ the ground-state wave function of nucleus $A$.  In LCA, the  complexity of the calculation is shifted from the wave functions to the operators. The complicated  $\left| \Psi_A \right>$ is obtained from a simple wave function $\left| \Phi_A \right>$ through the action of an operator
\begin{equation} \label{eq:corwav}
\left| \Psi _A \right> = 
\frac{1}
{\sqrt{\left<\Phi _A \right| \widehat{\mathcal{G}}^{\dagger} \widehat{\mathcal{G}} \left| \Phi _A \right>   }} \widehat{\mathcal{G}} \left|\Phi _A \right> \; ,
\end{equation}
where $\left|\Phi _A \right>$ is a Slater determinant wave function for nucleus $A$ and $\widehat{\mathcal{G}}$ is an operator that accounts for the SRC correlations. In LCA, we account for the central (Jastrow), tensor and spin-isospin SRC correlations  
\begin{eqnarray} \label{eq:SRCoperator}
\widehat{\mathcal{G}} & = &   \widehat{{\cal S}}  
\biggl( \prod _{i<j=1} ^{A} \biggl[ 1  
-  {g_c(r_{ij})} 
+  {f_{t\tau}(r_{ij}) } \widehat{S}_{ij} \vec{\tau}_i \cdot 
  \vec{\tau}_j \; 
\nonumber \\
& & + 
 { f_{\sigma \tau}(r_{ij}) } \vec{\sigma}_i \cdot \vec{\sigma}_j 
\vec{\tau}_i \cdot \vec{\tau}_j 
\biggr] \biggr)
\nonumber \\
& = &  
\widehat{{\cal S}}  
\biggl( \prod _{i<j=1} ^{A} \biggl[1 + \widehat{\mathcal{G}}_{ij} (r_{ij}) 
\biggr] \biggr)
\; ,
\end{eqnarray}
where $\widehat {{\cal S}}$ and $\widehat{S}_{ij}$ are the symmetrization and tensor operator.
In computing the $n^{A}(\vec{p})$ in LCA all terms are included up to order $\mathcal{O} (\mathcal{G} ^2)$ which implies that the impact of SRC on $n^A (\vec{p})$ is included as two-body operators. The   momentum distribution $n^{A}(\vec{p})$ is then the sum of two terms  (see also the diagrams (b) and (c) of Fig.~\ref{fig:schememomendis})
\begin{equation} \label{eq:momdis_has_twoterms}
    n^{A} (\vec{p}) = n^{A}_{\text{IPM}} (\vec{p}) + n^{A}_{\text{SRC}} (\vec{p}) + \mathcal{O} (\mathcal{G} ^3) \; .
\end{equation}
The first term $n^{A}_{\text{IPM}}$ (IPM stands for independent particle model) is reminiscent of independent nucleons 
\begin{equation} \label{eq:IPMcontri}
n^{A}_{\text{IPM}} (\vec{p})  \sim \sum_{N\in \{\text{p,n}\}} \sum_{\alpha} \left< N \alpha \right| a_{\vec{p}}^{\dagger} \; a_{\vec{p}}^{\phantom{\dagger}} \left| N \alpha \right>     \; ,
\end{equation}
where $\alpha$ extends over all occupied single-particle states in the Slater determinant 
$\left|\Phi _A \right>$. The second term is the result of the SRC operators of Eq.~(\ref{eq:SRCoperator}). It is determined by two-nucleon contributions of which the dominant contribution is of the form (see also diagram of Fig.~\ref{fig:schememomendis}(c))
\begin{widetext}
\begin{eqnarray} \label{eq:SRCcontri}
 n^{A}_{\text{SRC}}  (\vec{p}) & \sim &  \sum_{N N^{\prime} \in \{\text{p,n}\} } 
\sum_{\alpha \beta} 
\sum _{\vec{K} \; \vec{k} \; \vec{k} ^ {\prime}}
\widehat{\mathcal{G}}_{12} ^{\dagger} \left( \frac{\vec{K}}{2}+\vec{k}-\vec{p} \right)
\widehat{\mathcal{G}}_{12} ^{\phantom{\dagger}} \left( \frac{\vec{K}}{2}+\vec{k}^{\prime}-\vec{p} \right)
\nonumber \\
& & \times \left< N \alpha, N^{\prime} \beta \right| 
a_{\frac{\vec{K}}{2}+\vec{k}}^{\dagger} \;
a_{\frac{\vec{K}}{2}-\vec{k}}^{\dagger} \;
a_{\frac{\vec{K}}{2}+\vec{k}^{\prime}}^{\phantom{\dagger}} \;
a_{\frac{\vec{K}}{2}-\vec{k}^{\prime}}^{\phantom{\dagger}} \;
\left| N \alpha , N^{\prime} \beta \right>     \; .
\end{eqnarray}
\end{widetext}
For the sake of simplicity of the notation we make abstraction of the spin- and isospin dependence of the above two-body matrix elements. Note that in computing the $n^{A}_{\text{SRC}}  (\vec{p})$ one integrates over the center-of-mass momentum $\vec{K}$ of the correlated pair, as well as over the relative momenta  $\vec{k}$ and $\vec{k}^{\prime}$. \tcdr{The neglected terms of order $\mathcal{O} (\mathcal{G} ^3)$ in Eq.~(\ref{eq:momdis_has_twoterms}) include three-body correlations. The computation of those terms across the nuclear mass table is computationally prohibitive. There are indications, however, that the effect of three-nucleon correlations in the tail part of  the $n^{A} (\vec{p})$ is relatively small. Direct evidence comes from nuclear-matter calculations where the impact of three-nucleon effects has been studied~\cite{Rios:2013zqa}. Indirect theoretical evidence for the dominant role of two-nucleon correlations stems from the fact that quantum Monte Carlo calculations for light nuclei, that include all possible diagrams,  provide strong indications for $A$-to-$d$ scaling in the tail part of $n^{A} (\vec{p})$~\cite{Lonardoni:2017egu}.}

We restrict ourselves to spherically symmetric nuclei $n^{A}_{\text{SRC}}  (p)  \sim n^{A}_{\text{SRC}}  (\vec{p})$. Of high relevance for the isolation of the isospin composition of SRC is that both contributions (\ref{eq:IPMcontri}) and (\ref{eq:SRCcontri}) to $n^{A}(p)$ can be written as a sum of four terms  
\begin{equation} \label{eq:paircontributiontoNMD}
 n^{A}(p) \equiv \underbrace{n^{A}_{\text{pp}}(p) \; + \; n^{A}_{\text{pn}}(p)}_{n^{A}_{\text{p}}(p) \; (\text{proton part})} \; +  \; \underbrace{n^{A}_{\text{nn}}(p) \;  + \; n^{[1]}_{\text{np}}(p)}
 _{n^{A}_{\text{n}}(p) \; (\text{neutron part})}
 \; .
\end{equation}
 As schematically shown in Fig.~\ref{fig:schememomendis} the separation in the four pair combinations can done for both the IPM and the SRC contribution in Eq.~(\ref{eq:momdis_has_twoterms}). 
 \tcdr{For the SRC contribution of Eq.~(\ref{eq:SRCcontri}) the pairs $N \alpha, N^{\prime} \beta$ in the matrix elements give rise to the four pair combinations considered. In order to identify the pair combinations to the IPM contribution, one can rewrite Eq.~(\ref{eq:IPMcontri}) as 
 \begin{eqnarray}\label{eq:IPMcontri_re}
n^{A}_{\text{IPM}} (\vec{p})  & \sim & \sum_{N\in \{\text{p,n}\}} \sum_{N^{\prime} \in \{\text{p,n}\}} \sum_{\alpha} \sum_{\beta} \sum _{\vec{p}^{\prime}} 
\nonumber \\ 
& & \times \left< N \alpha \right| a_{\vec{p}}^{\dagger} \; a_{\vec{p}}^{\phantom{\dagger}} \left| N \alpha \right>       
\left< N^{\prime} \beta \right| a_{\vec{p}^{\prime}}^{\dagger} \; a_{\vec{p}^{\prime}}^{\phantom{\dagger}} \left| N^{\prime} \beta \right>
\; .
\end{eqnarray}
Herein, the summations $\sum_{N\in \{\text{p,n}\}} \sum_{N^{\prime} \in \{\text{p,n}\}}$ naturally give rise to four pair combinations.}
 Note that in identifying the different pair combinations contributing to $n^{A}(p)$ in the IPM, one integrates over the momentum of the second nucleon $N^{\prime}\beta$.   
We adopt the normalization convention $\int dp \; p^{2} n^{A}(p) =A$. In the adopted LCA, the four pair combinations stemming from pp, pn, nn and np contribute respectively a fraction $\frac{Z(Z-1)}{(A-1)}$, $\frac{NZ}{(A-1)}$, $\frac{N(N-1)}{(A-1)}$ and $\frac{NZ}{(A-1)}$ to the total norm $A$ of $n^{A}(p)$. These normalizations are not artificially imposed but obtained by expanding the matrix element of $\left<\Phi _A \right| \widehat{\mathcal{G}}^{\dagger} \widehat{\mathcal{G}} \left| \Phi _A \right>$ in the denominator 
 in Eq.~(\ref{eq:corwav}) up to the second order in the correlation operator~\cite{Vanhalst:2014cqa}. One finds  
 \begin{equation}
     \left<\Phi _A \right| \widehat{\mathcal{G}}^{\dagger} \widehat{\mathcal{G}} \left| \Phi _A \right> = A + \mathcal{C}_A +\mathcal{O} \left( \mathcal{G} ^3 \right) \; ,
 \end{equation}
 where $\mathcal{C}_A$ can be interpreted as a measure for the aggregated effect of SRC in nucleus A(N,Z)~\cite{Vanhalst:2014cqa}.  
 
Of great relevance for reactions involving nuclear targets is the probability distribution 
\begin{equation}
P^{A}(p) = p^2 n^{A}(p)/A \; \; \; \; \left( \int d p P^{A}(p) =1  \right)    
\end{equation}
to find a nucleon with momentum $p$ in A(N,Z). An immediate consequence of Eq.~(\ref{eq:paircontributiontoNMD}) is that  
\begin{equation} \label{eq:P(p)separation}
P^{A}(p)= 
\underbrace{P^{A}_{\text{pp}}(p) + P^{A}_{\text{pn}}(p)}_{P^{A}_{\text{p}}(p)\; (\text{proton part})}     
+ \underbrace{P^{A}_{\text{nn}}(p) + P^{A}_{\text{np}}(p)}_{P^{A}_{\text{n}}(p)\; (\text{neutron part})}\; .
\end{equation}
 
\begin{figure*}[htb]
\centering
 \includegraphics[width=0.90\columnwidth, viewport=46 78 534 411,clip]{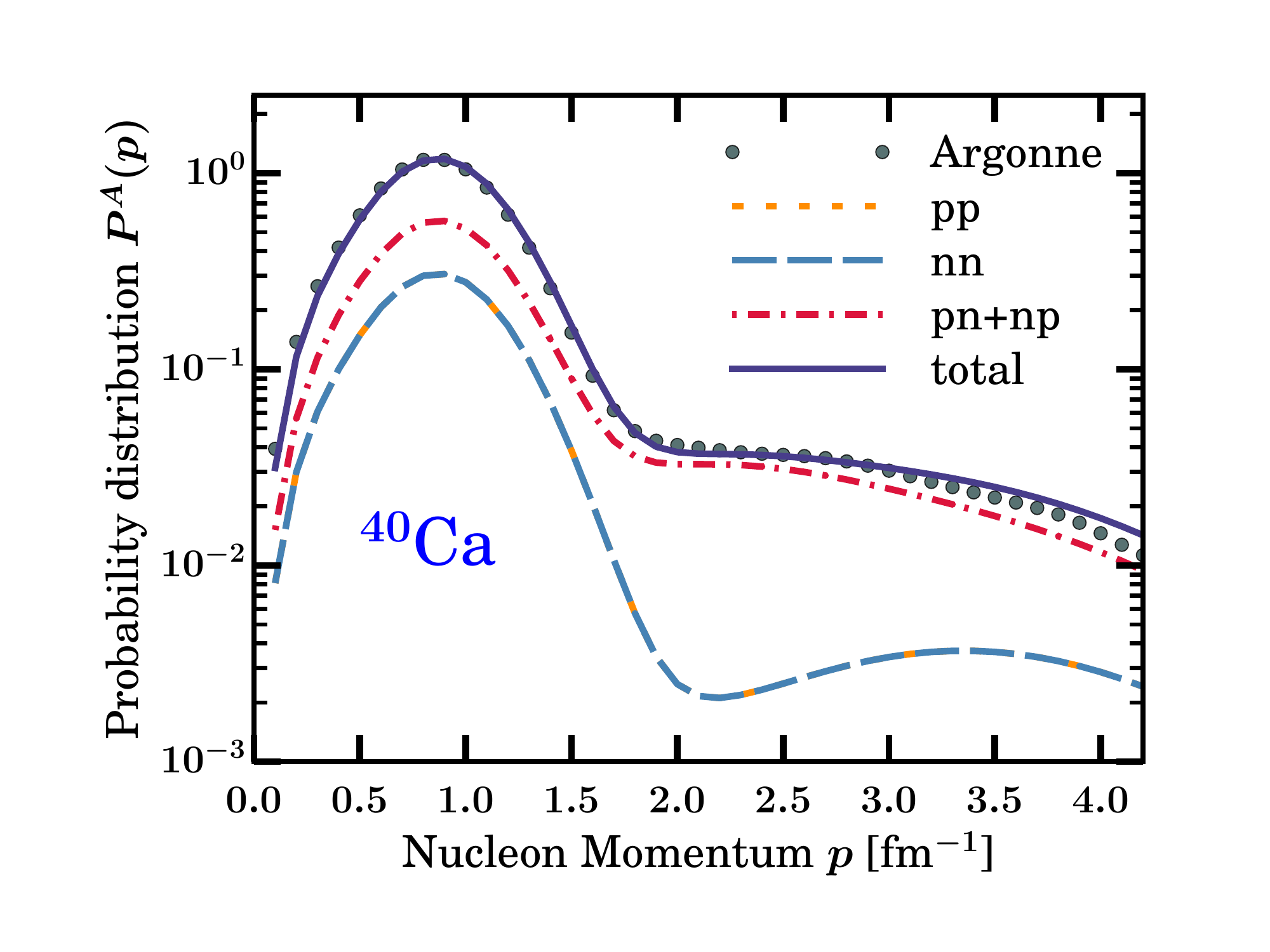}
 \includegraphics[width=0.8557416\columnwidth, viewport=70 78 534 411,clip]{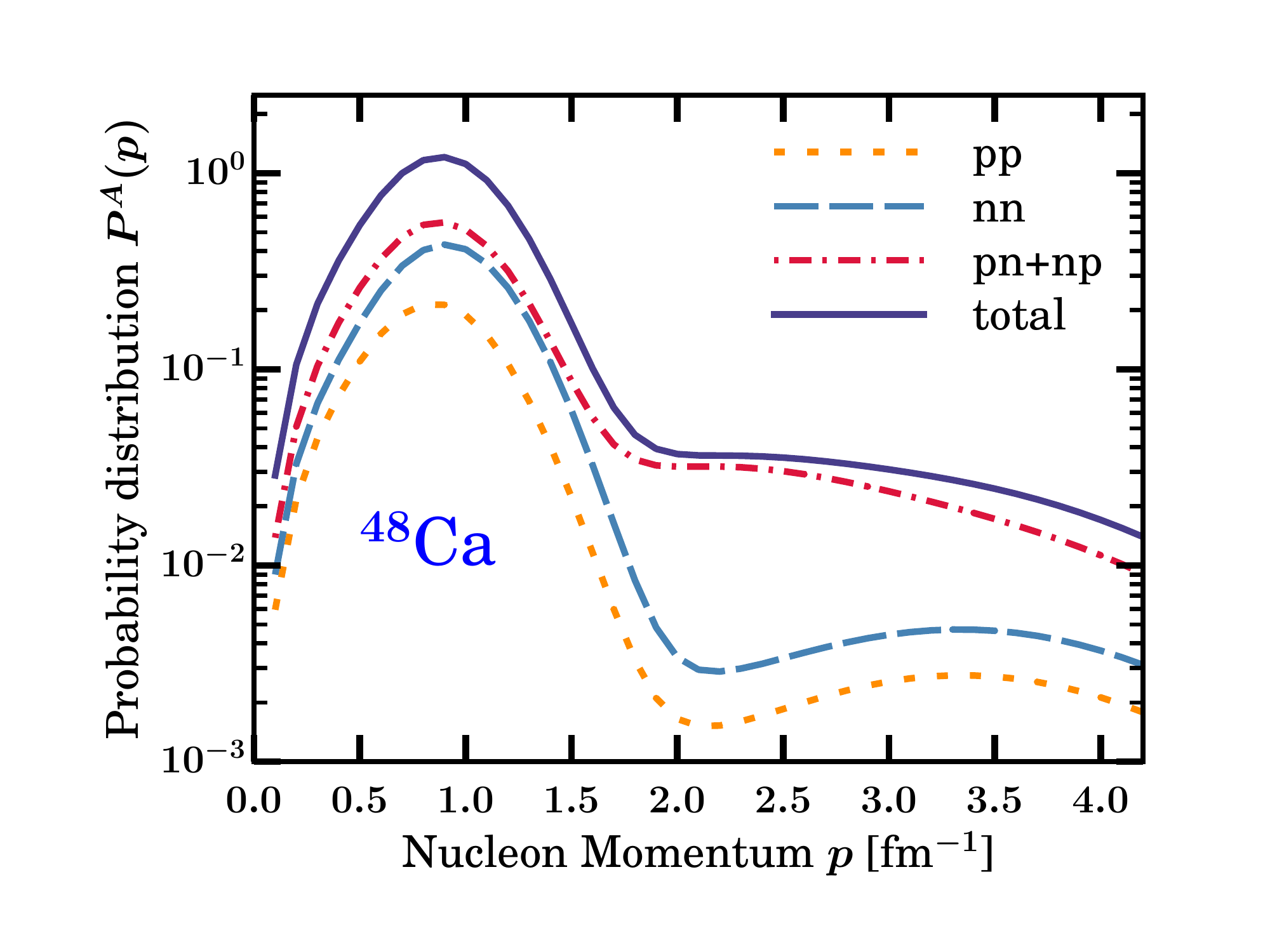}
\includegraphics[width=0.90\columnwidth, viewport=46 28 534 411,clip]{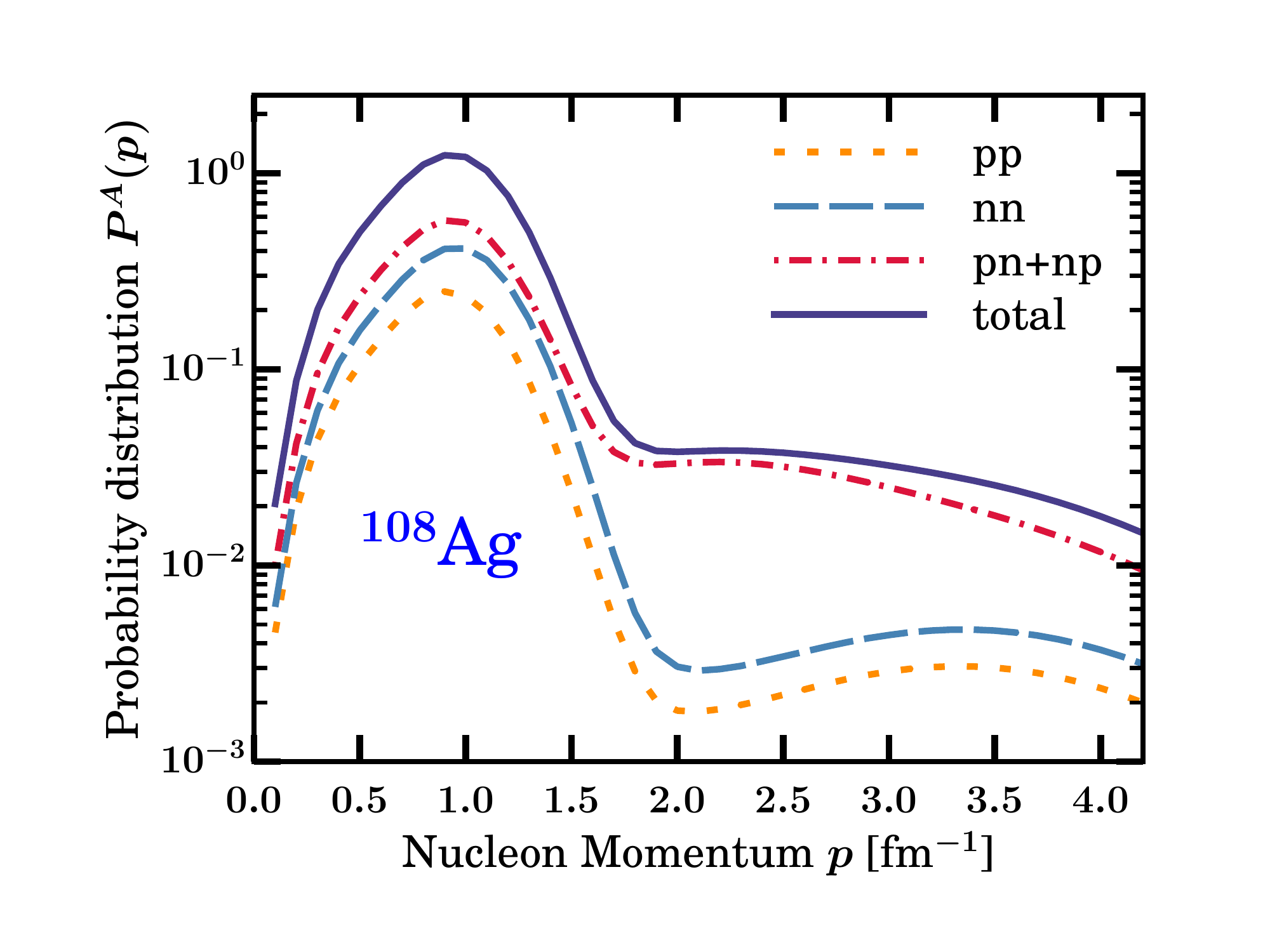}
 \includegraphics[width=0.8557416\columnwidth, viewport=70 28 534 411,clip]{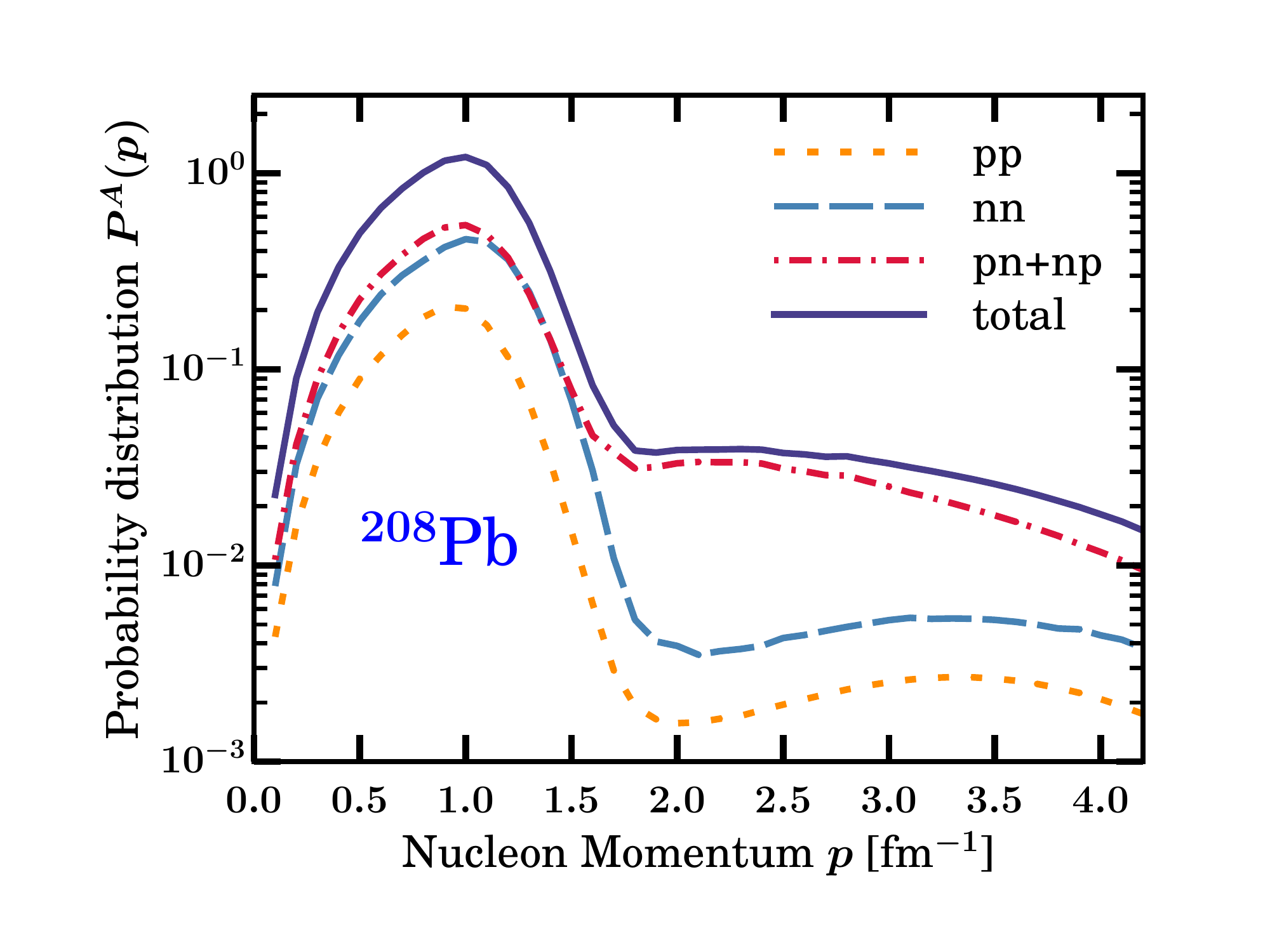}
\caption{The probability distribution $P^{A}(p)$ of finding a nucleon with momentum $p$ as computed in LCA for four nuclei. The separate contributions  from the four possible $NN^{\prime}$ combinations detailed in Eq.~(\ref{eq:P(p)separation}) are shown together with the total. For $^{40}$Ca we compare the LCA result for $P^A(p)$ with the one from the  Argonne group obtained with the effective AV18 nucleon-nucleon interaction~\cite{Lonardoni:2017egu}.} 
\label{fig:probabilitydistributions}
\end{figure*}

The correlation functions in Eq.~(\ref{eq:SRCoperator}) are input to the LCA approach. We adopt a data-driven methodology and use a set that has been systematically tested in comparisons of reaction-model calculations and SRC-driven data 
\cite{Blomqvist:1998gq, Onderwater:1998zz,Starink:2000qhh, Colle:2015ena}. The $f_{t 
\tau}(r_{12})$ and $f_{\sigma \tau}(r_{12})$ correlation functions are from a variational 
calculation~\cite{Pieper:1992gr}.  
An analysis of $^{12}$C$(e,e^{\prime}pp)$~\cite{Blomqvist:1998gq} and  $^{16}$O$(e,e^{\prime}pp)$~\cite{Starink:2000qhh} experimental results  systematically excluded ``soft'' central correlation functions $g_c$ and preferred  a ``hard'' $g_c(r_{12})$ inferred from a G-matrix calculation with the Reid soft-core interaction in nuclear matter~\cite{Dickhoff:2004xx}. 

Different interactions generate different correlations --particularly for the central ones-- and are sources of theoretical uncertainties in LCA. 
In Ref.~\cite{Ryckebusch:2018rct} we have presented LCA results with the ``hard'' $g_c$ from~\cite{Dickhoff:2004xx} and the ``soft'' $g_c$ from \cite{Pieper:1992gr} that is consistent with the adopted $f_{t 
\tau}$ and $f_{\sigma \tau}$.  The choice of the $g_c$ mainly affects the highest momentum parts of the single-nucleon momentum distributions. We found, however, that for light and medium-heavy nuclei the LCA in combination with the ``hard'' $g_c$ produces $n^{A}(\vec{p})$ that are in line with those from quantum Monte Carlo calculations with the effective AV18 $NN$ interaction. In addition, many extracted SRC properties are obtained from ratios of $n^{A}(\vec{p})$ for which the sensitivity to the choice of the $g_c$ is at the percent level~\cite{Ryckebusch:2018rct}.   

In this work, all calculations are performed in coordinate space with a ``hard'' $g_c$ and harmonic oscillator (HO) single-particle states $\left|N \alpha \right>$ as they offer the 
possibility to separate the pair's relative and center-of-mass~motions in  the pair wave functions $\left| N \alpha , N^{\prime} \beta \right>$ of Eq.~(\ref{eq:SRCcontri}) with the aid of 
Moshinsky brackets. As the major purpose of this study is to determine the systematic properties of the SRC scaling factors and their pair composition, we use the HO parameters from the global parametrization $\hbar \omega=45{A}^{-\frac {1} {3}} - 25{A}^{-\frac {2} {3}} $. More advanced calculations could find the optimum HO parameter for each specific nucleus but this complication is beyond the scope of the current paper. It has been numerically shown~\cite{ Vanhalst:2014cqa, Ryckebusch:1997gn, Colle:2015lyl,   Ryckebusch:1996wc, Colle:2013nna} and experimentally confirmed~\cite{Starink:2000qhh, Colle:2015ena} that the major source of SRC strength 
stems from correlation operators acting on IPM pairs in a nodeless relative $S$-state. This can be intuitively understood by noting that the 
probability of finding close-proximity IPM pairs is dominated by pairs in a nodeless 
relative $S$-state. Those $S$ wave functions are not very sensitive to the details of the mean-field potential which partially explains the robustness of the SRC properties in nuclei. 

Figure~\ref{fig:probabilitydistributions} displays the probability distributions $P^{A}(p)$ of Eq.~(\ref{eq:P(p)separation}) for four nuclei out of our  sample of fifteen nuclei.  In essence, there are two separated momentum regimes in the probability distributions. The underlying generative dynamics for the observed $p$ dependence has been discussed in great detail in Refs.~\cite{Vanhalst:2014cqa, Ryckebusch:2018rct}. Summarizing, the low-momentum part is reminiscent of the independent-particle model.  The high-momentum regime is characterized by a fat tail that displays a universal momentum dependence across the different nuclei.  The pair composition of the low-momentum regime is roughly determined by the combinatorics imposed by the neutron and the proton numbers $N$ and $Z$. In the SRC regime, on the other hand, there is an obvious proton-neutron dominance. As one approaches the highest momenta studied here, the proton-proton and neutron-neutron parts gain in relative importance relative to the dominant pn contribution. There is a degree of model-dependence in the probability distributions $P^{A}(p)$ of Fig.~\ref{fig:probabilitydistributions}~\cite{Bogner:2012zm,More:2017syr,Ryckebusch:2018rct}. Inclusion of SRC physics through the operators of Eq.~(\ref{eq:SRCoperator}) preserves the long-distance physics which makes the momentum dependence of the probability distributions below the Fermi momentum $p_F = 1.25$~fm$^{-1}$ independent of the choices made with regard to the correlation operators. For $p<p_F$ the major effect of the SRC correlations is to deplete  the $P^A(p<p_F)$ with a scaling factor that is model-dependent.    As we have shown in Ref.~\cite{Ryckebusch:2018rct} the high-momentum tail $P^A(p \gtrsim 2 ~\text{fm}^{-1} )$ displays some sensitivity to the choices made with respect to the correlation functions.   For $^{40}$Ca ---the heaviest nucleus for which ab-initio momentum distributions are available--- we can compare the LCA result for $P^{A}(p)$ with the one obtained with quantum Monte-Carlo methods using a realistic phenomenological NN interaction~\cite{Lonardoni:2017egu}. We observe a fair agreement providing confidence in our approach. In the following Sec.~\ref{subsec:tailpartofobmd} we connect the LCA predictions for the pair composition of the SRC tail of $P^{A}(p)$ to recent $A(e,e^{\prime}np)$ and $A(e,e^{\prime}pp)$ data.

\begin{figure*}[htb]
\centering
\includegraphics[width=0.80\columnwidth, viewport=46 28 534 411,clip]{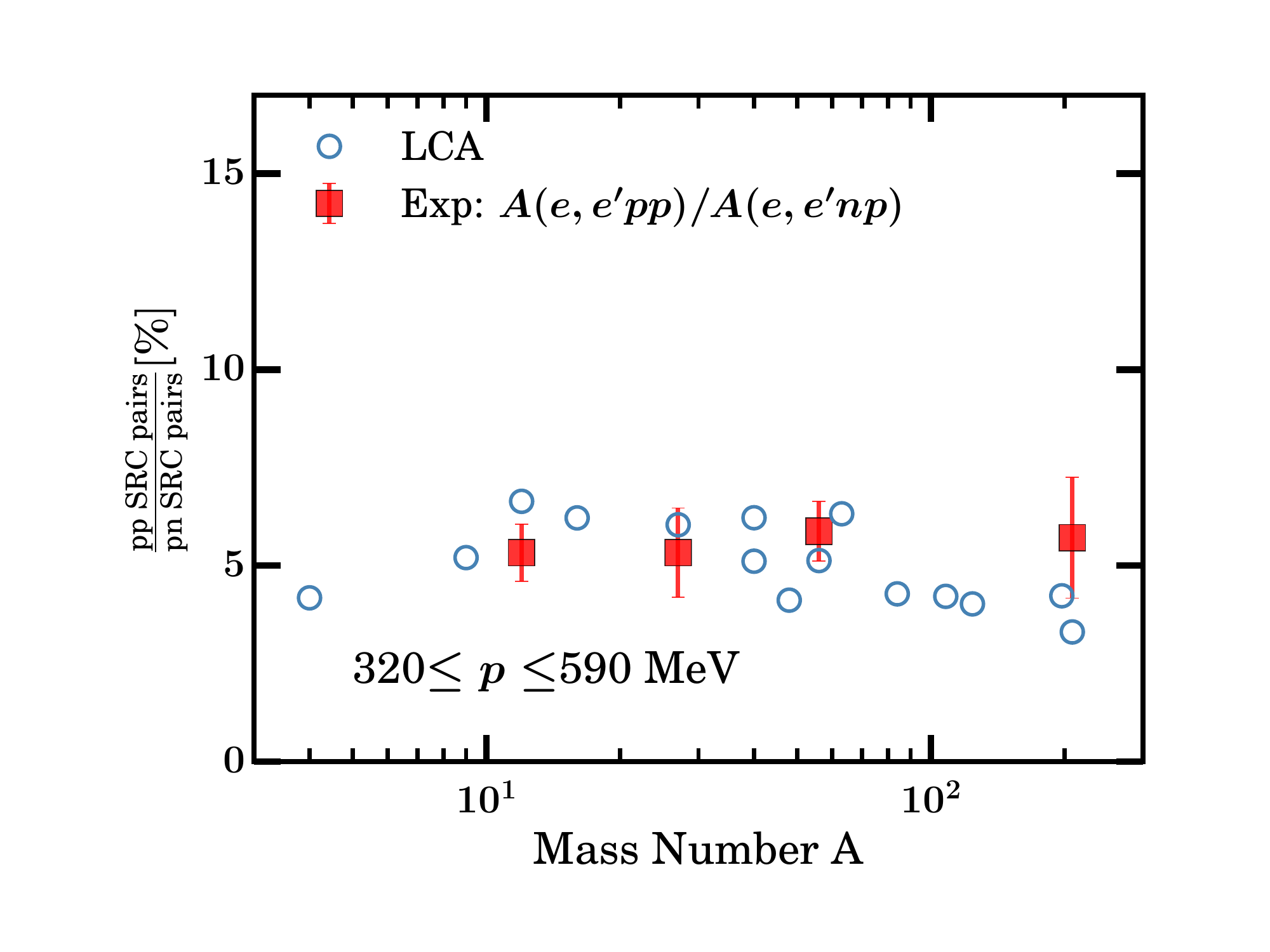}
 \includegraphics[width=0.721\columnwidth, viewport=94 28 534 411,clip]{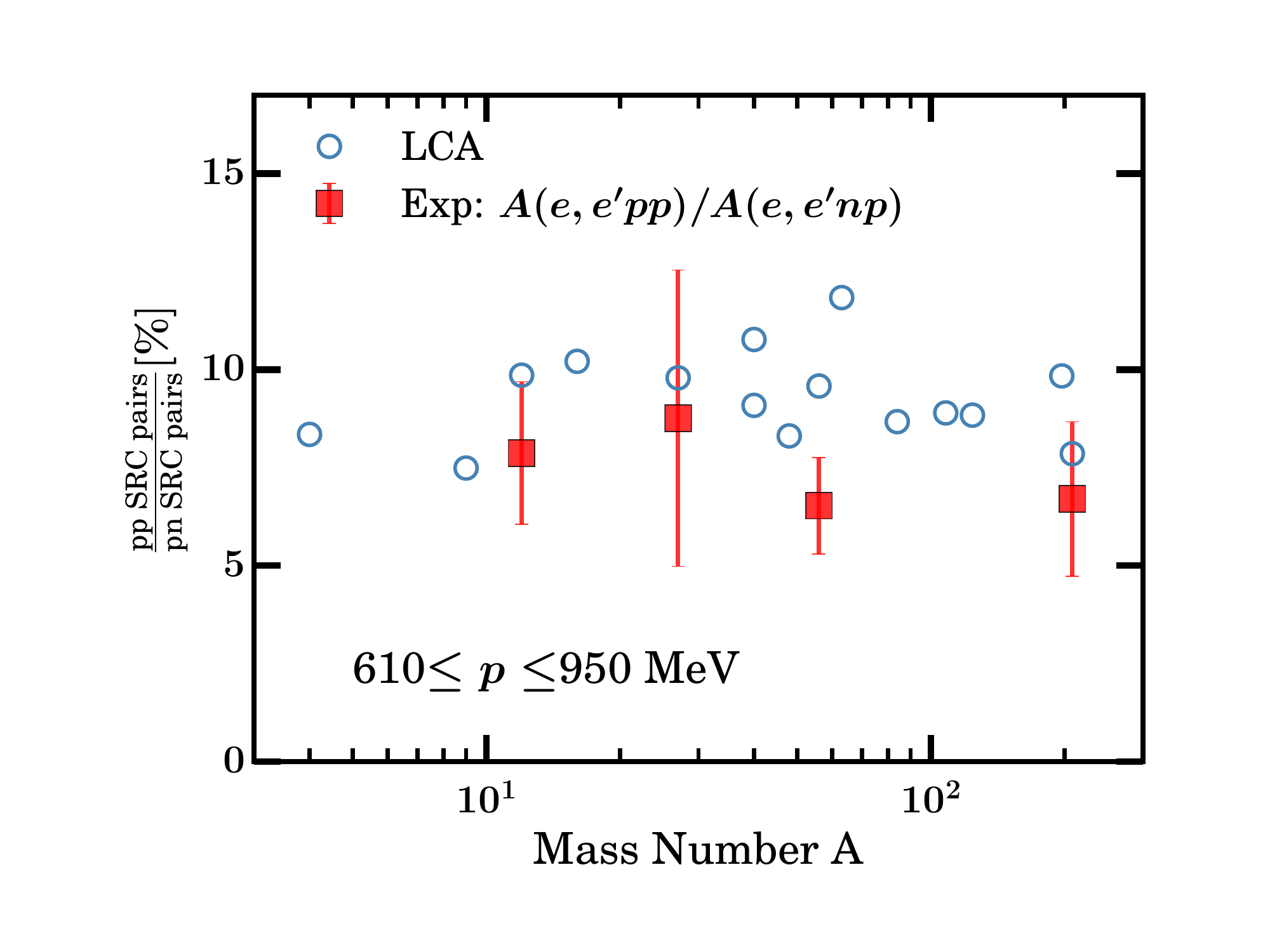}
\caption{The nuclear mass dependence of the ratios (in percent units) of pp-to-pn SRC correlated pairs in two momentum ranges above the Fermi momentum $p_F$. The blue circles are LCA predictions based on the ratio of the pp and pn contributions to the $P^A(p)$ in the selected momentum ranges. The data are from Ref.~\cite{Duer:2018sxh}. } 
\label{fig:eeppeepn}
\end{figure*}

\subsection{Tail part of single-nucleon momentum distributions}
\label{subsec:tailpartofobmd}
The relative weight of the pp and pn correlations in the tail of the momentum distribution can be ``measured'' by evaluating the ratio of the triple-coincidence $A(e,e^{\prime}pp)$ and  $A(e,e^{\prime}np)$ cross sections in a large-acceptance detector thereby probing
a large fraction of the phase space and imposing kinematical cuts selecting initial $2N$ SRC pairs~\cite{Duer:2018sxh}.  This amounts to evaluating a cross-section ratio of the form 
\begin{eqnarray}
& & \frac {\sigma_{en}}{2\sigma_{ep}}
\frac 
{ \sigma _{A}(e,e^{\prime}pp)}
 {  \sigma _{A}(e,e^{\prime}np)  } 
 \nonumber \\
 & & \sim 
 \frac{\text{Probability for pp SRC pair in A}}
      {\text{Probability for pn SRC pair in A}}
 \; . 
 \label{eq:neutrontoproton}
 \end{eqnarray}
 Here,  $\sigma_{ep}$ ($\sigma_{en}$) denotes the off-shell electron-proton (electron-neutron) cross section and $\sigma_A (e,e^{\prime}NN^{\prime})$ is the cross section for $NN^{\prime}$ knockout [where typically the ``fast'' nucleon has $p_N \gg p_F$ and the recoil nucleon of the SRC pair has $p_{N^\prime} =\mathcal{O}(p_F)$] aggregated over a certain initial-nucleon momentum range. The above ratio of probabilities can be made conditional on certain constraints for example with regard to the initial nucleon momenta where the picture is adopted that the ``fast'' nucleon $N$ has absorbed the virtual photon. As experiments probing SRC quantities often tag the momentum of the ``active'' nucleon and require events with an inactive A-2 core, the single-nucleon momentum distribution (see diagram (c) in Fig.~\ref{fig:schememomendis}) offers many opportunities for theory-experiment comparisons~\cite{Ryckebusch:2018rct}.   The theoretical counterpart of the ratio (\ref{eq:neutrontoproton}) reads
\begin{equation} \label{eq:thneutrontoproton}
\mathcal{C} \frac
 {\int_{p_{l}}^{p_{h}} dp P^{A}_{\text{pp}}(p)}
{\int_{p_{l}}^{p_{h}} dp \left[ P^{A}_{\text{np}}(p) + P^{A}_{\text{pn}}(p)  \right]} \; ,
\end{equation}
where $p_l$ and $p_h$ are determined by the experimental cuts for the inferred initial-nucleon momenta.    The ratio  of Eq.~(\ref{eq:neutrontoproton}) has recently been measured for carbon, aluminium, iron and lead for two initial momentum ranges $[p_l,p_h]$~\cite{Duer:2018sxh}. We use Eq.~(\ref{eq:thneutrontoproton}) to compare those data  to the LCA predictions. The results for the other 11 nuclei in our sample provide more detailed information about the variations across nuclei. The results of the theory-experiment comparisons are summarized in Fig.~\ref{fig:eeppeepn}. Apart from effects stemming from detector efficiencies for example, an important contribution to the factor $\mathcal{C}$ in the above equation are the final-state interactions (FSI). Attenuation will roughly equally affect protons and neutrons at the kinetic energies considered~\cite{Duer:2018sjb}. Single-charge exchange (SCX), however, is an important correction factor~\cite{Colle:2015lyl} when extracting SRC information for two-nucleon knockout reactions. Indeed a considerable amount of detected pp knockout events originate from virtual-photon absorption on pn SRC pairs. In Fig.~\ref{fig:eeppeepn} 
an overall reduction factor $\mathcal{C}=0.5$ in the theoretical ratio of Eq.~(\ref{eq:thneutrontoproton}) is used. In line with the data, the LCA predictions for the number of pp-to-pn SRC pairs are fairly constant among the fifteen nuclei in our sample and increase with increasing nucleon momentum. For a fixed momentum range, the variation in the predicted pp-to-pn SRC pair ratios across nuclei is of the order of few percent in line with the experimental observations.   
 
The experimentally determined $a_2^{exp}(A)$ of Eq.~(\ref{eq:a2meas}) is extracted from ratios of $A$-to-$d$ $(e,e^{\prime})$ cross sections. In the impulse approximation, those cross sections can be computed by integrating the phase-space weighted momentum distributions over selected ranges $\mathcal{R}$ determined by experimentally imposed conditions
\begin{equation} \label{eq:argument}
    a_2^{exp}(A) = \frac {2}{A} \frac
{\sigma^{A}(e,e^{\prime})}
{\sigma^{d}(e,e^{\prime})} \sim 
\frac {2}{A} 
\frac
{\int_{\mathcal{R}} dp p^2 n^{A}(p)}
{\int_{\mathcal{R}} dp p^2 n^{d}(p)}
\sim
\frac
{\int_{\mathcal{R}}  dp P^{A}(p)}
{\int_{\mathcal{R}}  dp P^{d}(p)} \; .
\end{equation}
Here, $\mathcal{R}$ is the momentum-range that corresponds to high initial-nucleon momenta and constant $A$-to-$d$ $(e,e^{\prime})$ cross sections. As those conditions can be associated with the tail part of the probability distributions of Fig.~\ref{fig:probabilitydistributions}, an estimate of the $ a_2^{exp}(A)$ of Eq.~(\ref{eq:argument}) can be obtained from the ratio of the weight in the tail parts of the computed $P^{A}(p)$
\begin{equation}
 \label{eq:a2fromSRCpart}
    a_2(A)  =   \frac
{\int_{p>2~\text{fm}^{-1}}  dp P^{A}(p)}
{\int_{p>2~\text{fm}^{-1}}  dp P^{d}(p)} \; .
\end{equation} 

\begin{figure}[htb]
\centering
\includegraphics[viewport=42 28 538 413, clip, width=0.95\columnwidth]{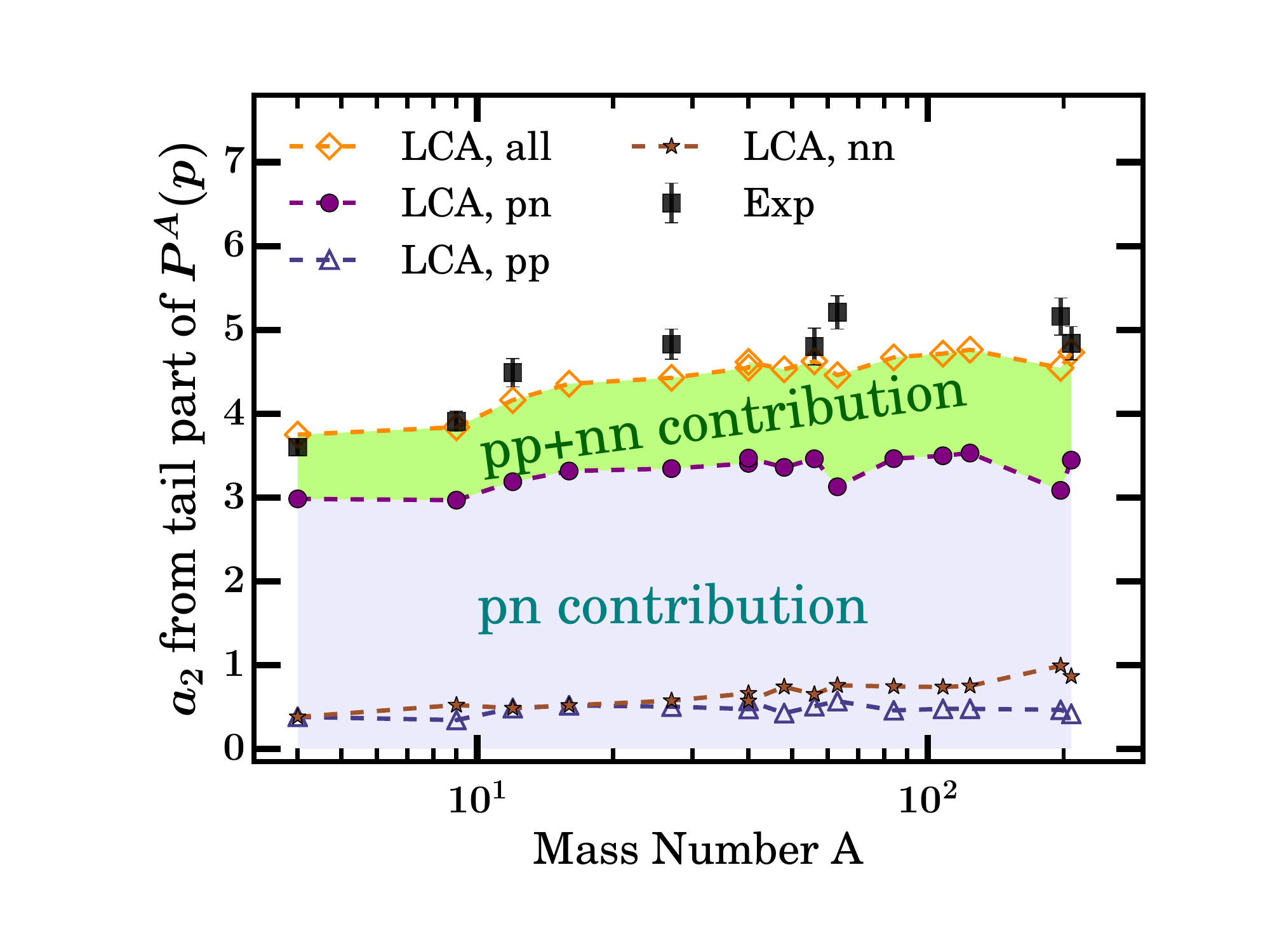}
\caption{LCA results for the  the SRC scaling factors $a_2(A)$ (orange open diamonds) along with the separate pp (blue open triangles), nn (brown stars) and pn (purple solid circles) contributions plotted versus atomic weight $A$. The shaded regions mark the pn (blue) and the pp+nn (green) contributions.  All results are computed from the $A$-to-$d$ weight of the tail part ($p>2~\text{fm}^{-1}$) of the nucleon probability distribution $P^A(p)$ [see Eq.~(\ref{eq:a2fromSRCpart})]. All $P^A(p)$ (including the deuteron one) are computed in LCA. The  $a_2^{exp}(A)$ data are from the extended data tables of Ref.~\cite{Schmookler:2019nvf} and include data from Ref.~\cite{Fomin:2011ng}.} 
\label{fig:a2_from_normalization}
\end{figure}

With the aid of the decomposition (\ref{eq:P(p)separation}) the contribution of the pp, nn and pn+np pairs to the numerator can be computed and the isospin composition of the SRC can be quantified.  The denominator of Eq.~(\ref{eq:a2fromSRCpart}) accounts for the weight of the tail part of the deuteron momentum distribution. Obviously, this number is model dependent \cite{More:2017syr,Marcucci:2018llz}.  For example, the denominator is 0.127 with the AV18 deuteron momentum distribution and is 0.103 with the LCA deuteron momentum distribution.  These numbers correspond with the tail part carrying about 10-13\% of the total probability in the deuteron. In Fig.~\ref{fig:a2_from_normalization} we present results of the above ratio for the 15 nuclei in our sample. For reasons of consistency, the deuteron probability distribution used is also computed in LCA.  From $^4$He to $^{208}$Pb the SRC scaling factor as computed with the aid of the Eq.~(\ref{eq:a2fromSRCpart}) has an increment of about 25\% --- from $\approx 3.8$ to $\approx4.8$ --- indicative of a very soft A-dependence that is also observed in the data. \tcdr{The soft A-dependence of the $a_2$ can be intuitively understood by considering that $a_2$ is determined by the local neighborhood of a nucleon. For light nuclei and increasing A$\lesssim 20$ the local neighborhood gradually fills up to reach approximate saturation for A$\gtrsim$20.} In the $A$-to-$d$ ratio the pn+np contribution is about 70\% of the total value and the pp part is of the order of 10\%. For $\frac{N}{Z}=1$ the pp and nn parts equally contribute. For $^{197}$Au and $^{208}$Pb, the two most asymmetric nuclei in our sample, the nn contribution approaches 20\% of the total. These numbers are to be compared to $\frac{N(N-1)}{A(A-1)}=0.36$ and 0.37, and are indicative of the isospin selectivity of the SRC.  In comparing the LCA predictions to the data it is important to realize that there are corrections to be applied, for example stemming from the center-of-mass motion of the NN pairs~\cite{Fomin:2011ng, Vanhalst:2012ur} and the fact that in a finite nucleus pairs can have excitation energies \cite{Janssen:1999xy}.  These corrections require either a full reaction model or a detailed Monte Carlo simulation and are outside the scope of this work.

\begin{figure*}[htb]
\centering
\includegraphics[width=0.75\columnwidth, viewport=46 28 534 411,clip]{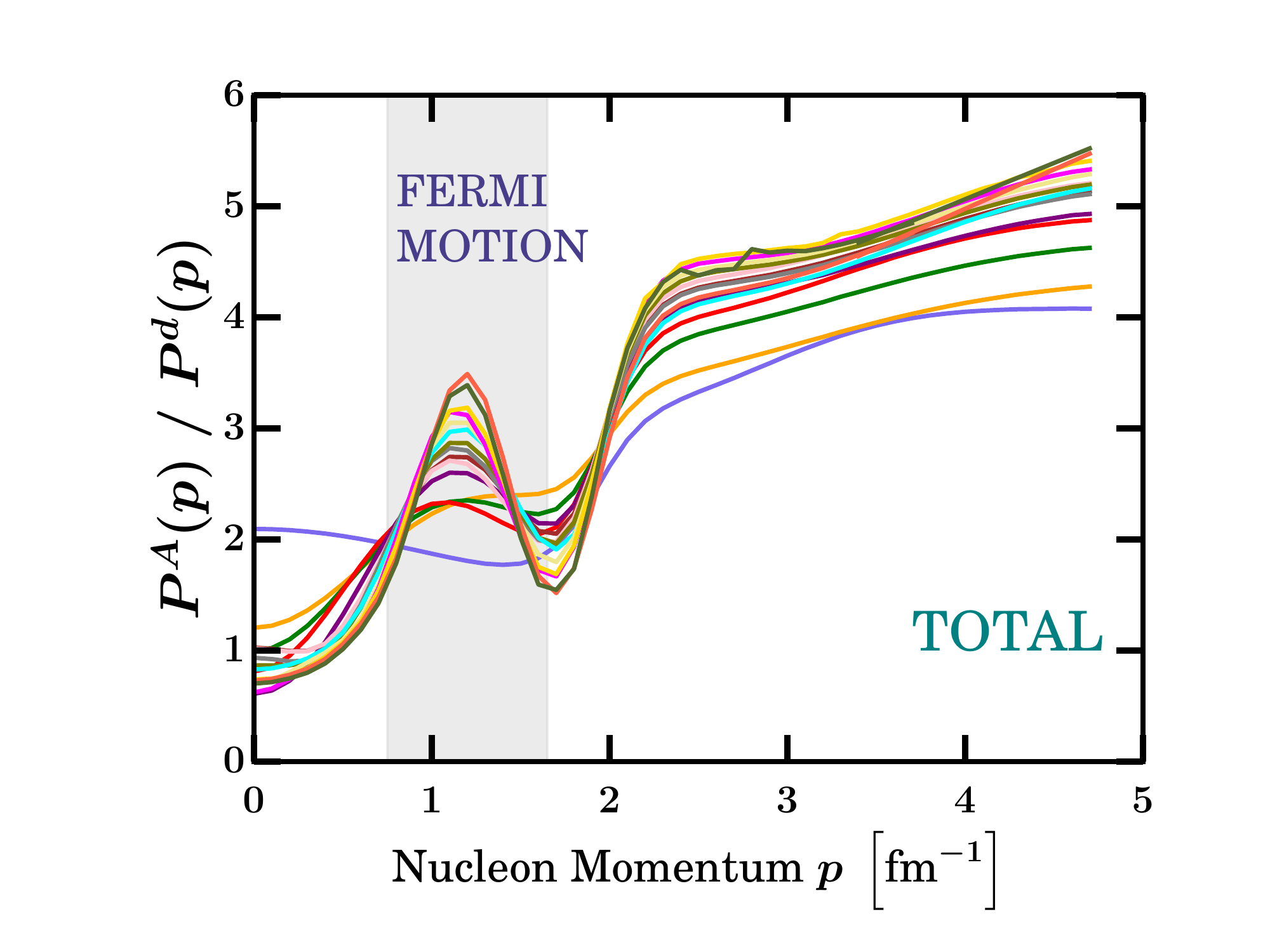}
 \includegraphics[width=0.75\columnwidth, viewport=46 28 534 411,clip]{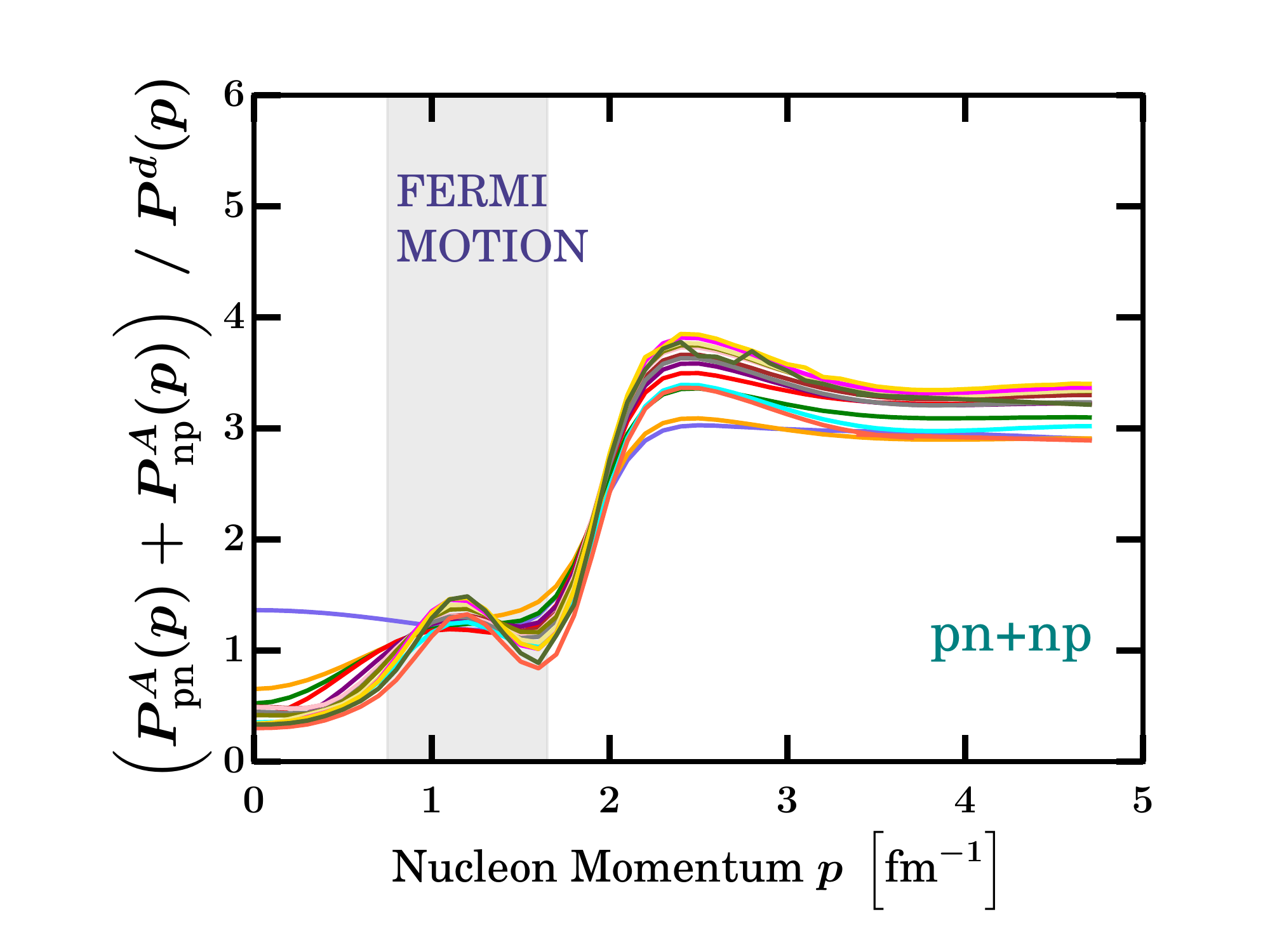}
\includegraphics[width=0.75\columnwidth, viewport=46 28 534 411,clip]{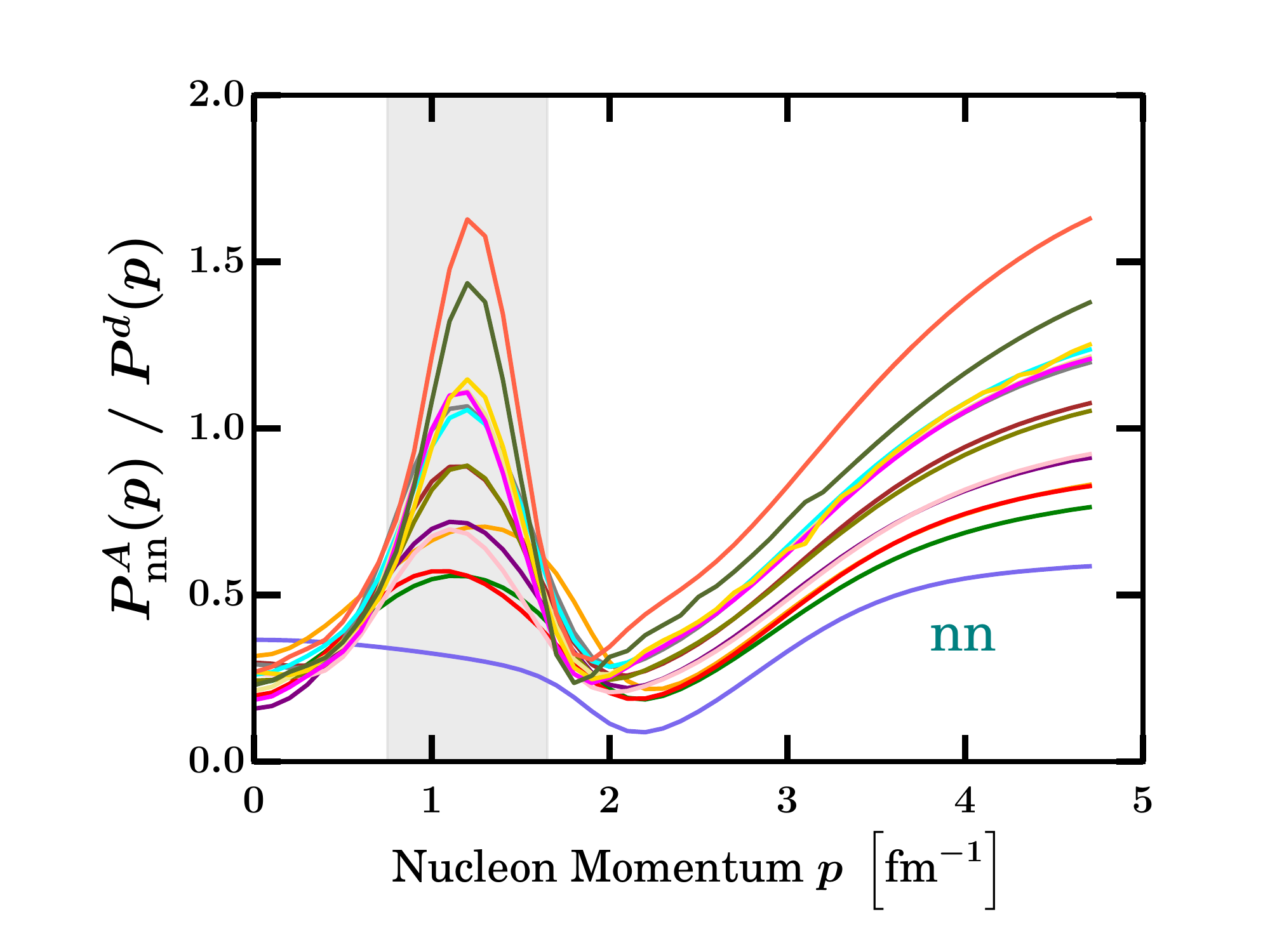}
 \includegraphics[width=0.75\columnwidth, viewport=46 28 534 411,clip]{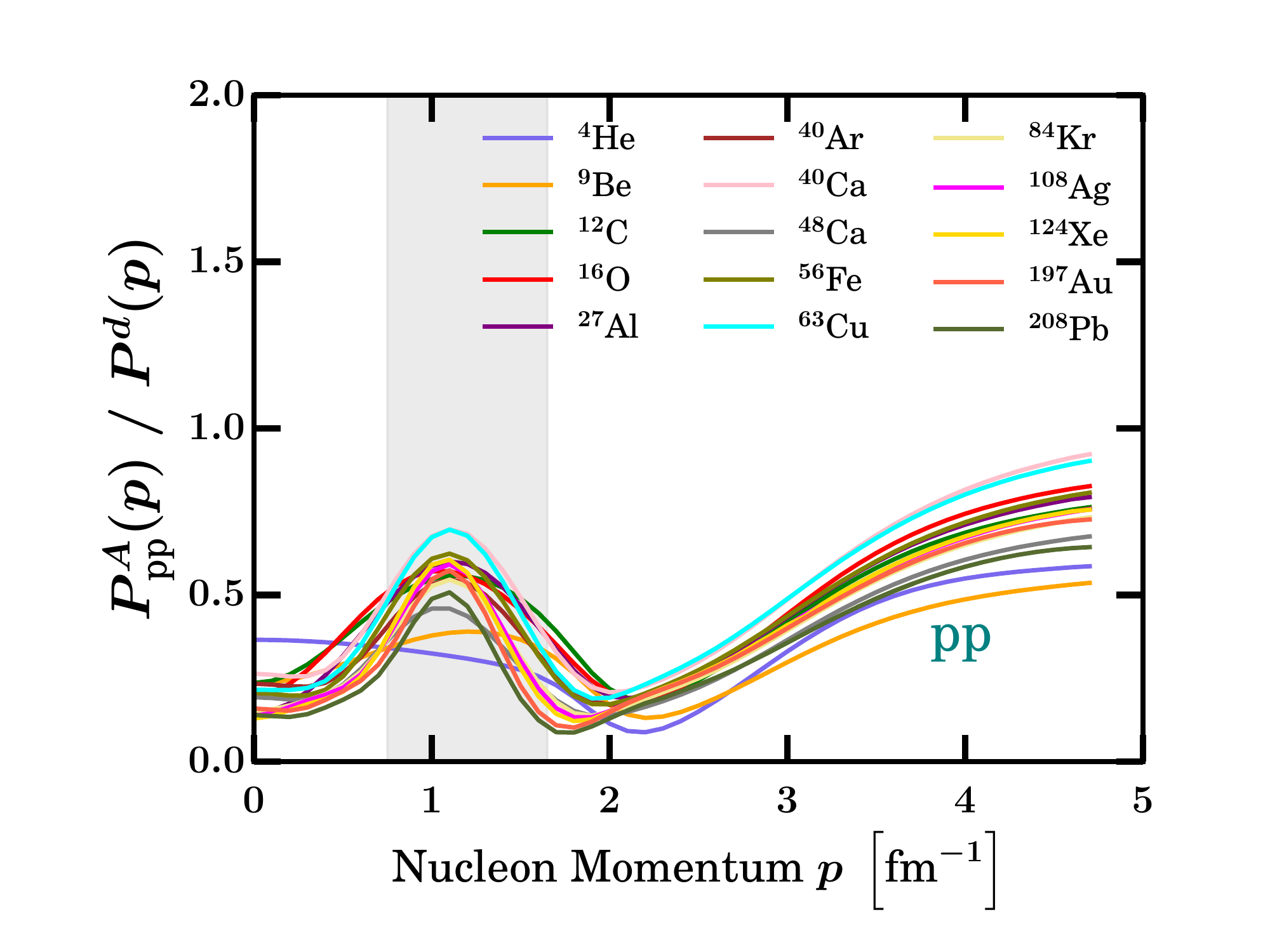}
\caption{The nucleon momentum dependence of the ratio of  the single-nucleon momentum distribution for nucleus A relative to the deuteron. Ratios are shown for the total momentum distribution, the sum of the proton-neutron and neutron-proton contributions, the neutron-neutron and proton-proton contributions. All the nucleon momentum distributions (including the deuteron one) are computed in LCA. Note that the y-axis scale for the nn and pp contributions (bottom figures) is different from the one used for the ``total'' and ``pn+pn'' ones (top figures). The grey shaded area shows the momentum range where Fermi motion dominates the displayed $A$-to-$d$ momentum distributions.} 
\label{figtailpartsLCA}
\end{figure*}
\subsection{Asymptotic single-nucleon momentum distributions and SRC scaling factors}
\label{subsec:asymptotics_and_SRCscaling}

\tcdr{In what follows we extract the SRC scaling factors $a_2$ from the asymptotic behavior of the $A$-to-$d$ momentum distributions.} The SRC scaling factors extracted from this method will be compared with those from Fig.~\ref{fig:a2_from_normalization} that use the aggregated weight in the SRC part of the $P^A(p)$. \tcdr{With this  comparison we explore the sensitivity of our methodology to the choices made with regard to the limits of integration in the tail parts of the single-nucleon momentum distributions.}       
We start with studying the momentum dependence of the $A$-to-$d$ probability distributions to identify the momentum ranges for which an $A$-to-$d$ scaling behavior emerges. 
In Fig.~\ref{figtailpartsLCA} we display the momentum dependence of the four ratios 
\begin{equation*} \label{eq:PpforAtodeuteron}
\frac{P^{A}(p)}{P^{d}(p)}, \; \;
\frac{ \left( P^{A}_{\text{pn}}(p) + P^{A}_{\text{np}}(p)\right) }{P^{d}(p)}, \; \;
\frac{P^{A}_{\text{nn}}(p)}{P^{d}(p)}, \; \; \frac{P^{A}_{\text{pp}}(p)}{P^{d}(p)} \;
\end{equation*}
for our sample of fifteen nuclei.  For each value of the momentum $p$, the $\frac{P^{A}(p)}{P^{d}(p)}$ provides the ratio of the per-nucleon probability of finding a nucleon in A(N,Z) relative to the deuteron. Obviously, any deviation from one is a measure of the medium dependence of the nucleon probabilities. The increased $A$-to-$d$ relative probability $\frac{P^A}{P^d}$ for $0.75 \lesssim p \lesssim 1.65$~fm$^{-1}$ is connected to Fermi motion in finite nuclei. In that momentum range the pp, nn  and pn contribute to $\frac{P^A}{P^d}$ roughly in accordance to their weight in $\frac{A(A-1)}{2}$. At $p \gtrsim 2.25$~fm$^{-1}$ one observes a plateau in the $\frac{P^A(p)}{P^d(p)}$ that extends to the highest momenta studied here. The plateau is characterized by an approximately universal momentum dependence of the $A$-to-$d$ ratio  $\frac{P^A(p)}{P^d(p)}$ at high momenta. Variations across nuclei can be captured by an SRC scaling factor that depends on A(N,Z). The onset of a plateau in the high-momenta results of $\frac{P^A(p)}{P^d(p)}$ in Fig.~\ref{figtailpartsLCA} provide support for the use of Eq.~(\ref{eq:a2fromasymptotics_p}) for extracting the SRC scaling factor. Indeed, in the limit of very high nucleon momenta relations between single-nucleon and two-nucleon momentum distributions can be established~\cite{Weiss:2015mba}.  The major trends in the $A$-to-$d$ ratios of Fig.~\ref{figtailpartsLCA} are in line with those of a study for six nuclei with $A\le 10$ reported in \cite{Weiss:2015mba}. The numerical calculations of that study also identify the $p \gtrsim 4$~fm$^{-1}$ region as the one suitable for extracting the $A$-to-$d$ SRC scaling factor.

 The SRC scaling factor extracted from the ``high-momentum'' ($p\approx 4.5$~fm$^{-1}$) behavior of $\frac{P^A(p)}{P^d(p)}$ for the total probability distribution for $^4$He is about 4. For the lightest nuclei in our sample --- $^{9}$Be, $^{12}$C, $^{16}$O,   $^{27}$Al  --- the SRC scaling factor increases with $A$ to reach the value of approximately 4.8 for $^{27}$Al. Small increments in the high-momentum  values of $\frac{P^A(p)}{P^d(p)}$ with increasing mass are observed for $A>27$.  
 
 A closer look at the pp, nn and pn+np contributions to the $\frac{P^A(p)}{P^d(p)}$ in  Fig.~\ref{figtailpartsLCA} indicates that the onset of the high-momentum plateau is most prominent for the pn+np parts. For the pp and nn contributions to $\frac{P^A(p)}{P^d(p)}$ the high-momentum scaling is not so pronounced as for the pn+np parts but there are indications that also these two ratios approximately saturate for $p \gtrsim 4~\text{fm}^{-1}$. After all, this is not so surprising given that there are no proton-proton and neutron-neutron correlations in the deuteron.
Along the same lines, there are stronger variations for the pp and nn contributions to  $\frac{P^A(p)}{P^d(p)}$ across nuclei than for the pn+np contribution. The variation in the pp and nn contribution to the $A$-to-$d$ SRC scaling factor cannot be captured by an A-dependence. In Sec.~\ref{subsec:SRCscalingfactor} it will be shown that the $N/Z$ ratio plays an important role in explaining those variations. For the pn+np parts the $A$-to-$d$ SRC corrections at nucleon momenta $p \gtrsim 4$~fm$^{-1}$ are substantially larger than the corrections attributed to the Fermi motion in the $0.75 \lesssim p \lesssim 1.65$~fm$^{-1}$ range. For the pp and nn parts, on the other hand, the SRC $A$-to-$d$ modifications are of the same order as the ones attributed to Fermi motion.  \tcdr{Following up on the above discussion about the fact that the $ \frac{ P^{A}_{\text{pp}}(p)}{P^{d}(p)}$ and $\frac{P^{A}_{\text{nn}}(p) }{P^{d}(p)}$ plateaus are not really very flat at high nucleon momentum, we have confirmed that the corresponding $ \frac{ P^{A}_{\text{pp}}(p)}{P^{^{4}\text{He}}(p)}$ and $\frac{P^{A}_{\text{nn}}(p) }{P^{^{4}\text{He}}(p)}$ plateaus are far more flat.}

The results of Fig.~\ref{figtailpartsLCA} 
provide support for extracting to $A$-to-$d$ SRC scaling factor from the high-momentum behavior of the ratio 
\begin{equation} \label{eq:a2fromasymptotics}
a_2 (A) = \lim _{\text{high} \; p} \frac{P^{A}(p)}{P^{d}(p)} \;  ,
\end{equation}
where ``high $p$'' stands for the momentum range for which a plateau in the $A$-to-$d$ momentum distribution is visible. We stress that the above expression for $a_2(A)$ is similar in vein to the ones of Eqs.~(\ref{eq:a2fromasymptotics_r}) and (\ref{eq:a2fromasymptotics_p}) that have  been derived within the context of effective field theories (EFTs) \cite{Chen:2016bde, Lynn:2019vwp} and of the contact formalism \cite{Alvioli:2016wwp, Weiss:2016obx,Cruz-Torres:2019fum}. The following results are computed in the spirit of the expression (\ref{eq:a2fromasymptotics}) whereby we have defined ``high p'' as being well into the asymptotic $p$ region of the $A$-to-$d$ momentum distributions (see~Fig.~\ref{figtailpartsLCA}). As is clear from Fig.~\ref{figtailpartsLCA} the plateau in the $A$-to-$d$ probability distributions can be clearly identified for the dominant pn contribution. For the pp and nn parts, on the other hand, the $A$-to-$d$ scaling is approximately realized but at the highest $p$ there are indications of saturation. Therefore, the ``high-$p$ limit'' $\lim _{\text{high}\; p}$ of the $A$-to-$d$ probability distributions  of Eq.~(\ref{eq:a2fromasymptotics}) was numerically evaluated by means of the ratios
\begin{equation} \label{eq:highpeval}
a_2 (A) = \lim _{\text{high}  \; p} \frac{P^{A}(p)}{P^{d}(p)} \approx
\frac{\int  _{\Delta p^{\text{high}}} \; dp \; P^A(p)}
{\int _{\Delta p^{\text{high}}} dp P^d(p)}    \; 
\end{equation}
Similar expressions are used to evaluate the $\lim _{\text{high}\; \; p}$ of the ratios $\frac{P^A_{pp}}{P^d}$, $\frac{P^A_{nn}}{P^d}$, $\frac{P^A_{np}}{P^d}$, and $\frac{P^A_{pn}}{P^d}$.  Based on the location of the occurrence of the plateaus in Fig.~\ref{figtailpartsLCA} the range $\Delta p^{\text{high}}$ of $\text{high}-p$ values is  $\left[p^{\text{high}}_l,p^{\text{high}}_u\right]$ with $p^{\text{high}}_l > 3.8$~fm$^{-1}$ and $p^{\text{high}}_u < 4.5$~fm$^{-1}$.  In the process of selecting the boundaries of the range $\Delta p^{\text{high}}$ we have also taken into consideration the good practice of keeping the nucleon momenta smaller than the nucleon mass in non-relativistic calculations. The range $ \Delta p^{\text{high}} $ is of the order of 100~MeV. There are some systematic uncertainties in our approach that are connected with the selection of the ``high-momentum'' regime. Referring to the results of Fig.~\ref{figtailpartsLCA} the highest uncertainty stems from the nn contribution. A very conservative estimate of the error induced by defining a ``high-p'' regime of the ratios of $A$-to-$d$ is that it induces an uncertainty in the extracted $a_2^p$ and $a_2^n$ of the order of 0.50, which corresponds to an error of about 10\%. 

\begin{figure}[htb]
\centering
\includegraphics[viewport=42 28 538 413, clip, width=0.95\columnwidth]{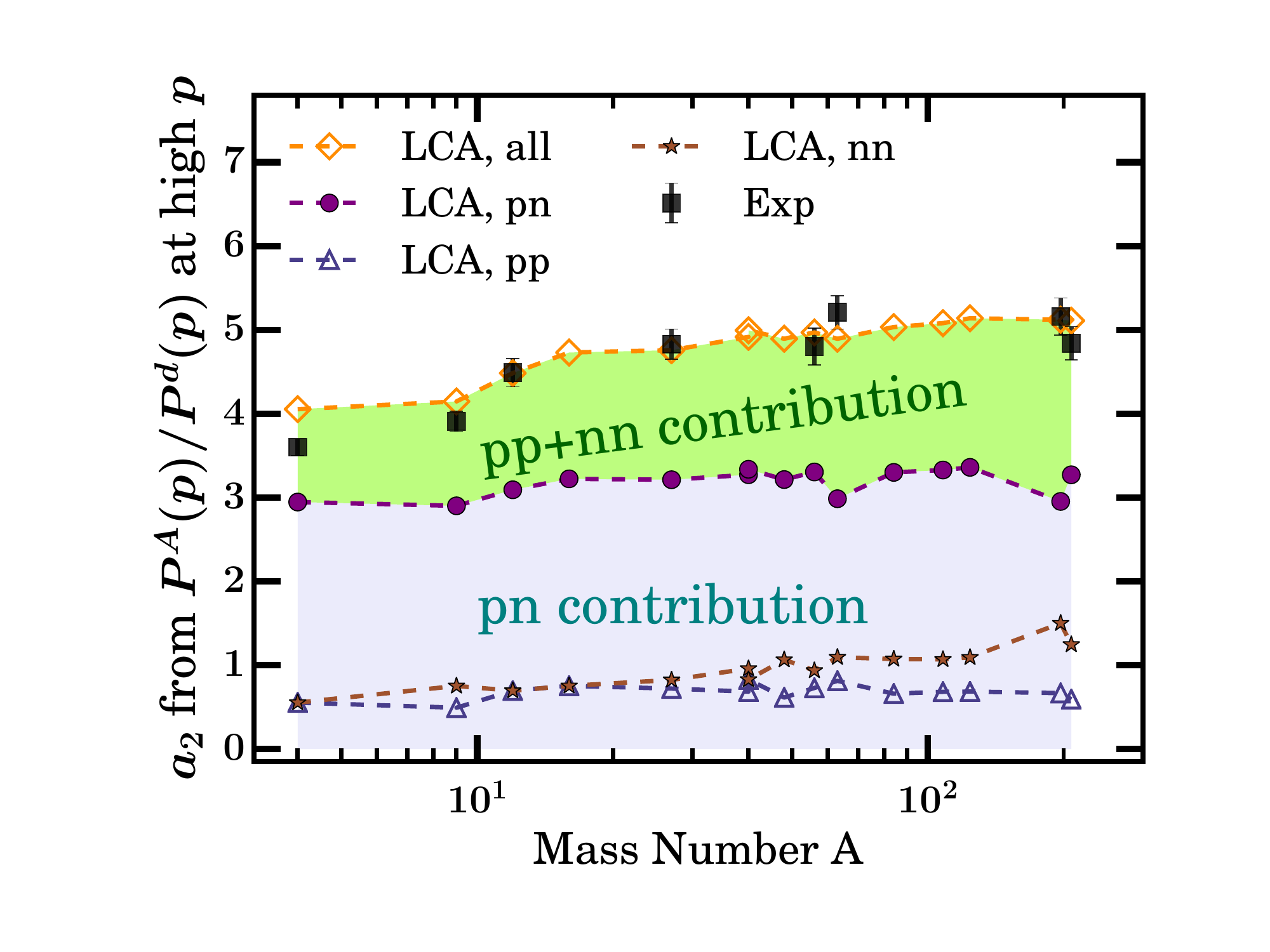}
\caption{LCA results for the  the SRC scaling factors $a_2(A)$ (orange open diamonds) along with the separate pp (blue open triangles), nn (brown stars) and pn (purple solid circles) contributions plotted versus atomic weight $A$. The shaded regions mark the pn (blue) and the pp+nn (green) contributions. All results are obtained from the asymptotic high-$p$ behavior of the $A$-to-$d$ single-nucleon nucleon momentum distribution $P^A(p)$ (see Eq.~(\ref{eq:a2fromasymptotics}) and text for details). All $P^A(p)$ (including the deuteron one) are computed in LCA. The  $a_2^{exp}(A)$ data are from the extended data tables of Ref.~\cite{Schmookler:2019nvf} and include data from Ref.~\cite{Fomin:2011ng}.} 
\label{fig:a2_from_normalization_scaling}
\end{figure}

\begin{table*}
\caption{\label{tab:all_results} Results for the proton and neutron SRC scaling factors $a_2^p$ and $a_2^n$ and related quantities as computed in LCA for a sample of 15 nuclei.  The $a_2^p$ and $a_2^n$ are computed with the aid of the Eq.~(\ref{eq:a2p_and_a2n}). The $- \frac {d R_{EMC}} {dx}$ are computed with the aid of Eq.~(\ref{eq:EMCwithsandv}).}
\begin{ruledtabular}
\begin{tabular}{lcccccccc}
Nucleus & $\frac{N}{Z}$ & $a_2^p$  & $a_2^n$  & $\frac{Z a_2^p + N a_2^n}{A}$ & $a_2^{exp}$ & $\frac{Z a_2^p - N a_2^n}{A} $ & $- \frac {d R_{EMC}} {dx}$ & $ b_2^{exp}$     \\ \hline \\
$^4$He  &  1.00  &  4.05  &  4.05  &  4.05  &  $3.60 \pm 0.10$ \cite{Hen:2012fm} &  -0.00  &  0.268  &  $0.207 \pm 0.025$ \cite{Seely:2009gt}, $0.222 \pm 0.045$ \cite{PhysRevD.49.4348} \\ 
$^9$Be  &  1.25  &  4.37  &  3.97  &  4.15  &  $3.91 \pm 0.12$ \cite{Hen:2012fm}  &  -0.26  &  0.336  &   $0.326 \pm 0.026$ \cite{Seely:2009gt}, $0.283 \pm 0.028$ \cite{PhysRevD.49.4348} \\ 
$^{12}$C  &  1.00  &  4.48  &  4.48  &  4.48  & $4.75 \pm 0.16$ \cite{Hen:2012fm}  &  -0.00  &  0.306  & $0.340\pm 0.022$ \cite{Schmookler:2019nvf},   $0.285 \pm 0.026$ \cite{Seely:2009gt},    \\
  &    &    &    &    & $4.49\pm 0.17$ \cite{Schmookler:2019nvf} &    &    &  $0.322 \pm 0.033$ \cite{PhysRevD.49.4348}  \\
$^{16}$O  &  1.00  &  4.73  &  4.73  &  4.73  &   &  -0.00  &  0.328  &   \\ 
$^{27}$Al  &  1.08  &  4.83  &  4.69  &  4.76  & $4.83\pm 0.18$  \cite{Schmookler:2019nvf} &  -0.10  &  0.354  &   $0.347\pm 0.022$ \cite{Schmookler:2019nvf} \\ 
$^{40}$Ar  &  1.22  &  5.15  &  4.72  &  4.92  &   &  -0.28  &  0.408  &   \\ 
$^{40}$Ca  &  1.00  &  4.99  &  4.99  &  4.99  &   &  -0.00  &  0.351  &  \\ 
$^{48}$Ca  &  1.40  &  5.33  &  4.59  &  4.89  &   &  -0.46  &  0.446  &   \\ 
$^{56}$Fe  &  1.15  &  5.13  &  4.83  &  4.97  & $4.80\pm 0.22$ \cite{Schmookler:2019nvf} &  -0.21  &  0.397  &  $0.472 \pm 0.023$ \cite{Schmookler:2019nvf}, $0.391 \pm 0.025$ \cite{PhysRevD.49.4348}  \\ 
$^{63}$Cu  &  1.17  &  5.01  &  4.80  &  4.89  & $5.21 \pm 0.20$ \cite{Hen:2012fm}   &  -0.29  &  0.407  & $0.391 \pm 0.025$ \cite{PhysRevD.49.4348}   \\ 
$^{84}$Kr  &  1.33  &  5.38  &  4.77  &  5.03  &   &  -0.42  &  0.450  &   \\ 
$^{108}$Ag  &  1.30  &  5.38  &  4.85  &  5.08  &   &  -0.39  &  0.449  &   \\ 
$^{124}$Xe  &  1.30  &  5.42  &  4.92  &  5.14  &   &  -0.41  &  0.458  &   \\ 
$^{197}$Au  &  1.49  &  5.34  &  4.98  &  5.12  & $5.16 \pm 0.22$ \cite{Hen:2012fm}  &  -0.84  &  0.554  &  $0.511 \pm 0.030$ \cite{PhysRevD.49.4348} \\ 
$^{208}$Pb  &  1.54  &  5.64  &  4.77  &  5.11  & $4.84\pm 0.20$  \cite{Schmookler:2019nvf} &  -0.66  &  0.513  & $0.539 \pm 0.020$ \cite{Schmookler:2019nvf}   \\
\end{tabular}
\end{ruledtabular}
\end{table*}

Figure~\ref{fig:a2_from_normalization_scaling} shows the SRC scaling factors as computed with the aid of the expression~(\ref{eq:highpeval}) for the fifteen nuclei in our sample. We provide the total SRC scaling factor as well as the separated proton-proton, neutron-neutron and proton-neutron contributions to it. The results can be compared with those of Fig.~\ref{fig:a2_from_normalization} that use the aggregated weight in the SRC part of the momentum distributions [Eq.~(\ref{eq:a2fromSRCpart})]. All-in-all the two methods to determine the SRC scaling factors $a_2(A)$ provide comparable predictions for all nuclei considered in this work. The method based on the evaluation of the high-p limit of $\frac{P^A(p)}{P^d(p)}$ tends to predict larger pp and nn contributions to  $a_2(A)$.  
\tcdr{In essence, Figs.~\ref{fig:a2_from_normalization} and \ref{fig:a2_from_normalization_scaling} use different ranges of integration ($p>2$~fm$^{-1}$ and $3.8<p<4.5$~fm$^{-1}$) to determine the $a_2(A)$. The difference between the extracted numbers can be seen as a measure for the sensitivity to the adopted ranges of integration in the tail parts of the single-nucleon momentum distributions.}
\begin{figure}[htb]
\centering
\includegraphics[viewport=42 28 538 413, clip, width=0.95\columnwidth]{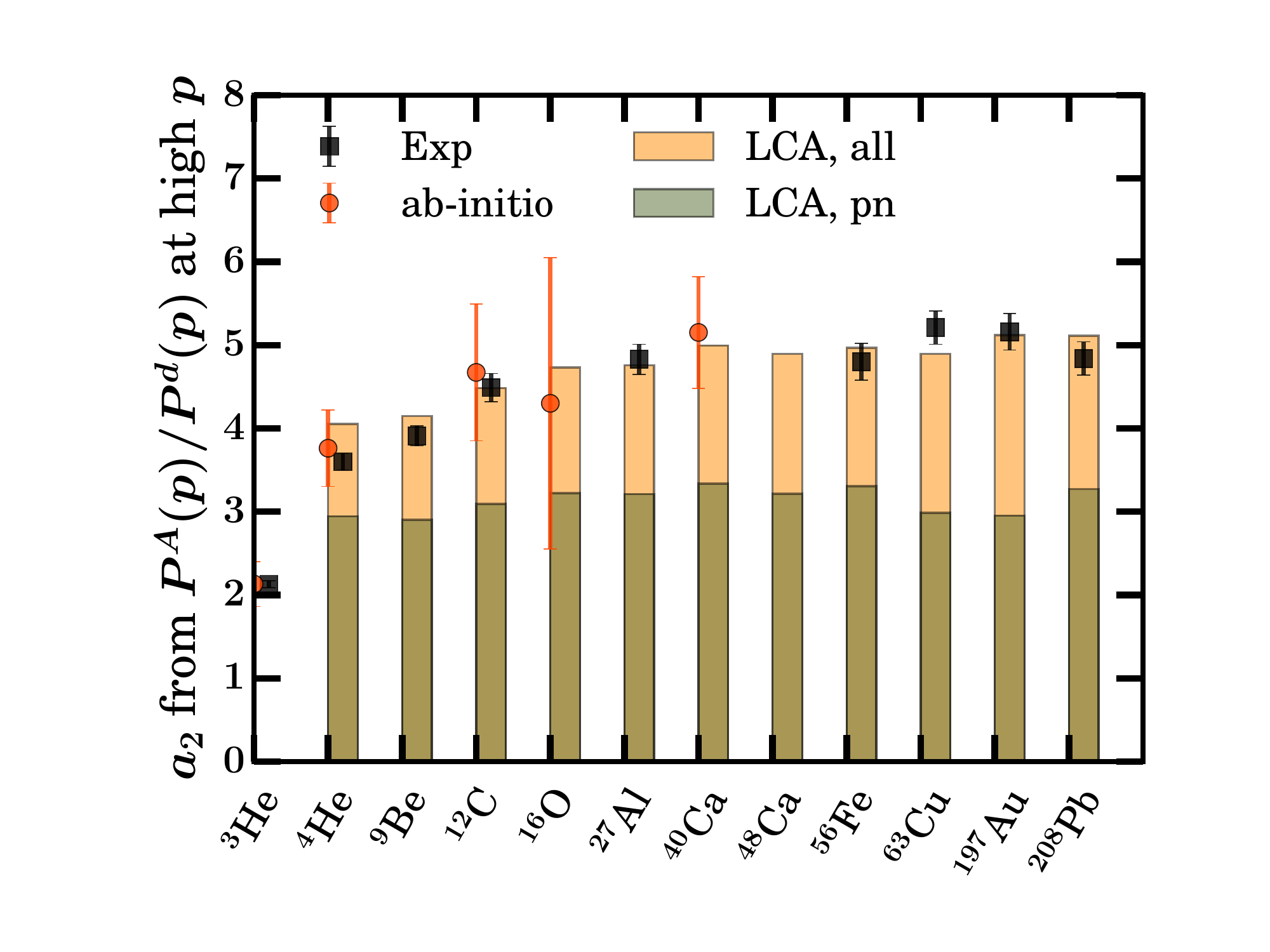}
\caption{Comparison of the LCA results for the SRC scaling factors $a_2(A)$ with those from ab-initio calculations and with measured values. The ab-initio results are from Table~II of Ref.~\cite{Lynn:2019vwp} (results referred to as ``N$^2$LO E$\tau$ $R_0$=1.0~fm'' except for $^{40}$Ca  where the sole available ab-initio result uses the ``AV4$^{\prime}$+UIX$_c$'' nucleon-nucleon interaction). The  $a_2^{exp}(A)$ data are from the extended data tables of Ref.~\cite{Schmookler:2019nvf} and include data from Ref.~\cite{Fomin:2011ng}.} 
\label{fig:a2_th_exp}
\end{figure}

Figure~\ref{fig:a2_th_exp} compares the LCA results for $a_2$ using Eq.~(\ref{eq:a2fromasymptotics}) with those from ab-initio calculations and data. Within the error bars the LCA results are compatible with those from ab-initio calculations and the data. For $^{40}$Ca and $^{48}$Ca we predict $a_2=4.99$ and $a_2=4.89$. The heaviest nucleus for which ab-initio results are available is $^{40}$Ca and the LCA result of 4.99 compares well with the ab-initio result of 5.15$\pm$0.67. The LCA prediction for $^{12}$C ($a_2=4.48$) is in line with the measured values: $a_2^{exp}=4.49 \pm 0.17$~\cite{Schmookler:2019nvf} and $a_2^{exp}=4.75 \pm 0.16$~\cite{Hen:2012fm}.

\subsection{Proton and neutron SRC scaling factors for symmetric and asymmetric nuclei}
\label{subsec:SRCscalingfactor}

In an asymmetric $N \ne Z$ nuclear environment one can anticipate that the $A$-to-$d$ SRC scaling factors are different for protons and neutrons~\cite{Ryckebusch:2018rct}. 
In order to quantify this and gain better insight into the $\frac{N}{Z}$ dependence of the SRC scaling factors  we introduce 
\begin{eqnarray} \label{eq:a2panda2n}
a_2 (A) & = &  \lim_ {\text{high\;} p} \frac{P^{A}(p)}{P^{d}(p)}  =    \lim_ {\text{high\;} p} \frac{P^{A}_{\text{p}}(p)+ P^{A}_{\text{n}}(p)  }{P^{d}(p)} 
\nonumber \\
&\equiv & \frac {Z a_2^{p} (A) + N a_2^{n} (A)}{A} \; .
\end{eqnarray}
We remind that with the adopted normalization conventions one has 
\begin{equation}
\int dp  P^{A}_{\text{p}}(p)  =  \frac{Z}{A}\; , \; \; \; \;
\int dp  P^{A}_{\text{n}}(p)  =  \frac{N}{A} \; ,  
\end{equation}
and that for the deuteron we can formally write $P^d = P^d_p + P^d_n = 2 P^d _p = 2 P^d _n $. Rearranging the above equations leads to the definitions   
$a_2^{p} (A)$ and $a_2^{n} (A)$ 
\begin{equation} \label{eq:a2p_and_a2n}
a_2^{p} (A) = \lim_ {\text{high\;} p} \frac{A \; P_p^A} {Z \; P_p^d} \; , \; \; \; \;
a_2^{n} (A) = \lim_ {\text{high\;} p} \frac{A \; P_n^A} {N \; P_n^d} \; .
\end{equation}
Accordingly, $a_2^{p} (A)$ encodes the per-proton probability to find a high-momentum proton in A(N,Z) relative to $d$. Similarly,  $a_2^{n} (A)$ encodes the per-neutron probability to find a high-momentum neutron in A(N,Z) relative to $d$. Note that $a_2^{p} (A=d) = a_2^{n} (A=d) =1$. A deviation from $a_{2}^p (A) =a_{2}^{n} (A)$ is reminiscent of the fact that there are differences in the per-proton and per-neutron dynamical modifications attributed to the short-distance structure of the nuclear environment in A(N,Z).  We stress that in the absence of pp and nn correlations the high-momentum tails of $P_p^A$ ($P_n^A$) would only receive a $P_{pn}^A$ ($P_{np}^A$) contribution. For predominant proton-neutron correlations one has that $P_{pn}^A(p > p_F) \approx P_{np}^A(p > p_F)$ and one can infer that  
\begin{equation} \label{eq:limitpndominant}
a_2^p(A) \approx \frac{N}{Z}a_2^n(A) \; \qquad \text{(for pn exclusivity)} \; .  \end{equation}
Note that the $a_2^p$ and $a_2^n$ defined in Ref.~\cite{Schmookler:2019nvf} obey this ``pn-exclusivity'' inspired relation by construction and that in the limit of vanishing pp and nn correlations one has the strict relationship between $a_2$, $a_2^p$ and $a_2^n$
\begin{equation} \label{eq:a2pnstrictrelation}
a_2 (A) = \frac {2 Z} {A} a_2^p(A) = \frac {2 N} {A}   a_2^n(A) \; \; \text{(for pn exclusivity)} \; .  
\end{equation}

\begin{figure}[htb]
\centering
\includegraphics[viewport=42 28 538 413, clip, width=0.95\columnwidth]{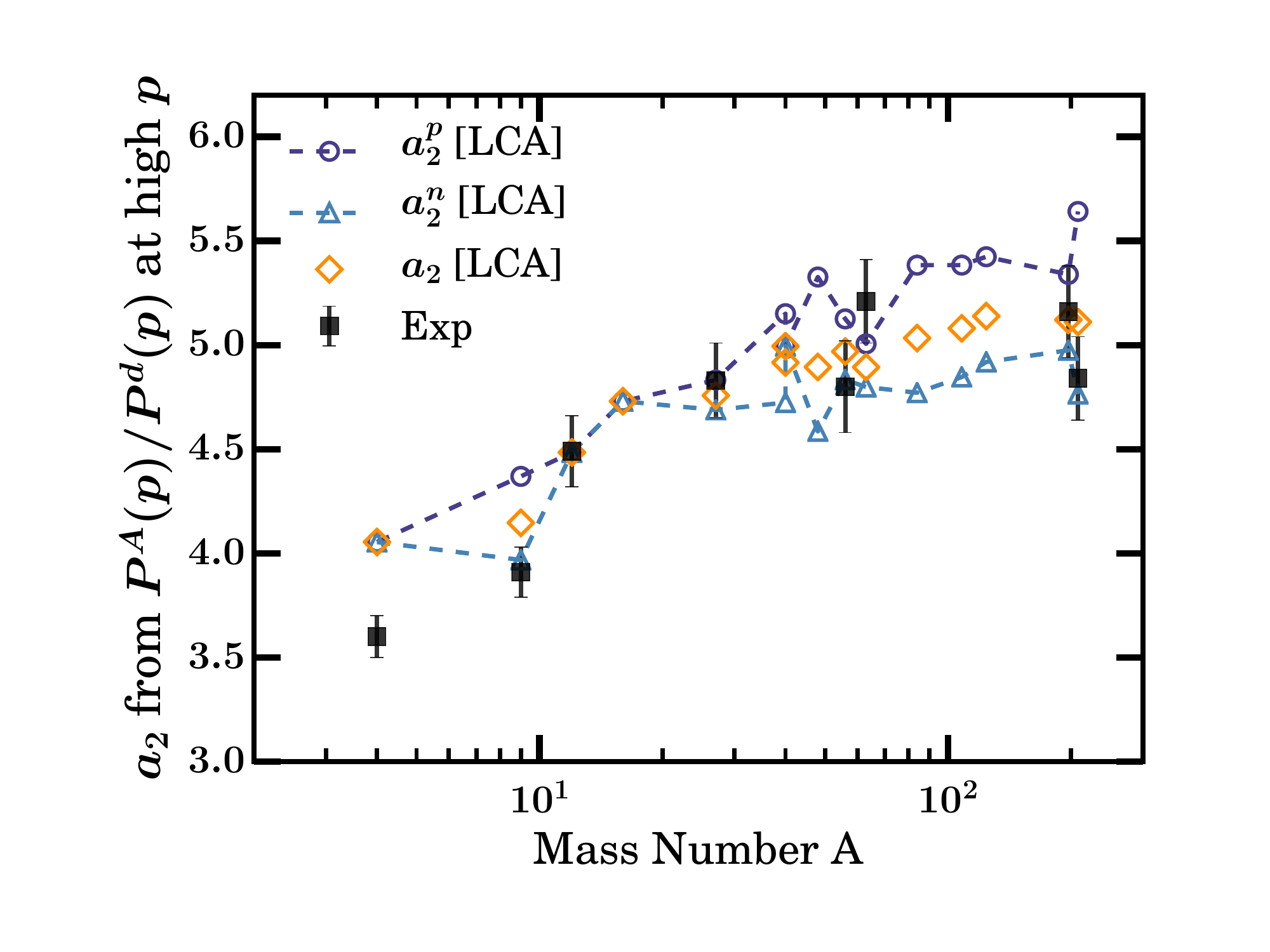}
\includegraphics[viewport=42 28 538 413, clip, width=0.95\columnwidth]{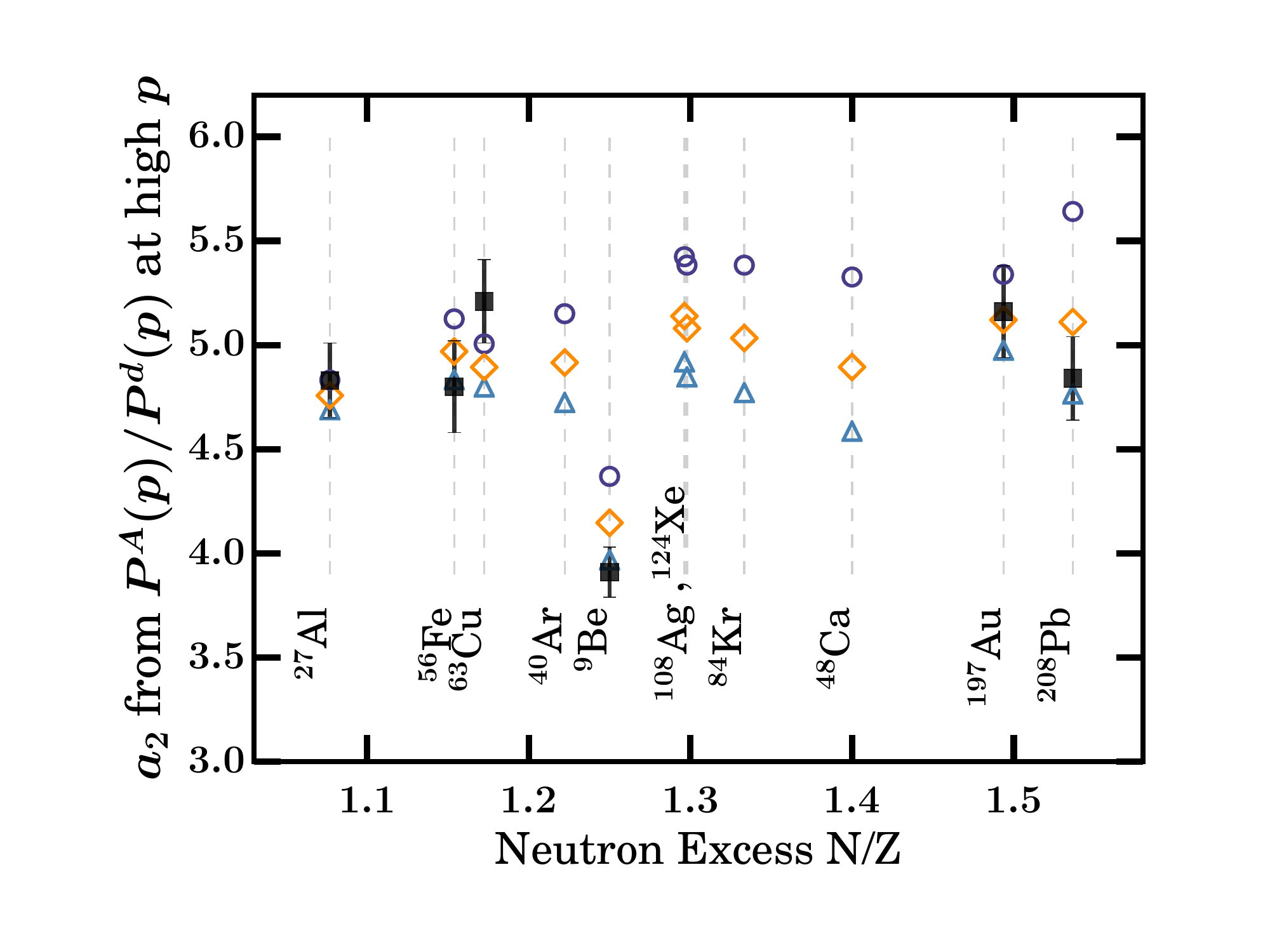}
\caption{The LCA predictions for the SRC scaling factors  for a sample of 15 nuclei as a function of $A$ (top panel) and $N/Z$ (bottom panel). Results are displayed for the proton $a_2^p$ (circles), neutron $a_2^n$ (triangles) and total $a_2$ (diamonds) SRC scaling factors. The LCA results are obtained with the expression (\ref{eq:highpeval}) with $\Delta p^{\text{high}} \equiv \left[3.90~\text{fm}^{-1}, 4.40 ~\text{fm}^{-1} \right] $. The experimental data are from the extended data tables of Ref.~\cite{Schmookler:2019nvf} and include data from Ref.~\cite{Fomin:2011ng}.} 
\label{fig:a2results}
\end{figure}


In Fig.~\ref{fig:a2results} we show the evolution of the computed $a_2^{p} (A)$, $a_2^{n} (A)$ and $a_2(A)$ with mass number $A$ and proton-to-neutron ratio $\frac{N}{Z}$. The $\lim_{\text{high}\;p}$ in Eq.~(\ref{eq:a2p_and_a2n}) is numerically evaluated as outlined in Eq.~(\ref{eq:highpeval}) and the discussion following this expression.   All numerical values for the  $a_2^{p} (A)$ and $ a_2^{n}(A)$ are also contained in Table~\ref{tab:all_results}. 
For asymmetric nuclei $\frac{N}{Z}>1$ one finds that $a_2^{p} (A) > a_2^{n}(A)$. This implies that per nucleon the proton minority component contributes more to the SRC scaling factors than the neutron majority component, which is in line with previous observations~\cite{PhysRevC.79.064308, Sargsian:2012sm,Ryckebusch:2018rct}. This result is not surprising given that pn exclusivity gives rise to the relation $a_2^p \approx \frac{N}{Z} a_2^n$ [Eq.~(\ref{eq:limitpndominant})]. The ratio $\frac{a_2^p(A)}{a_2^n(A)}$ increases with growing $\frac{N}{Z}$. Whereas for $A\gtrsim 27$ the   
$a_2^{n}(A)$ varies between 4.59 and 4.98, one observes that the $a_2^{p}(A)$ increases with $\frac{N}{Z}$ to reach a value of about 5.5 for the most neutron-rich nuclei considered here ($^{197}$Au and $^{208}$Pb). For $^{48}$Ca ($\frac{N}{Z}=1.4$) the 
$a_2^{p}(A)$ is about 15\% larger than the $a_2^{n}(A)$. The outlier in the SRC scaling factors at $\frac{N}{Z}=1.25$ is for $^{9}$Be and reflects the fact that for light nuclei the $a_2$ is about one unit smaller than for medium-heavy and heavy nuclei. In other words, $^9$Be is the sole light asymmetric nucleus in our sample. Table~\ref{tab:all_results} also lists the values of $\frac{Za_2^p - N a_2^n}{A}$  a quantity that vanishes for $N=Z$ nuclei. In asymmetric nuclei, $\frac{Za_2^p - N a_2^n}{A}$ approaches zero in the scenario of prevailing proton-neutron correlations. Obviously, $\frac{Za_2^p - N a_2^n}{A}$ grows increasingly negative with $\frac{N}{Z}$ which reflects the fact that the nn correlations  increase in importance with growing $\frac{N}{Z}$ as can be inferred from Figs.~\ref{figtailpartsLCA} and \ref{fig:a2_from_normalization_scaling}.  


\begin{figure}[ht]
%
\centering
\includegraphics[viewport=42 28 538 413, clip, width=0.95\columnwidth]{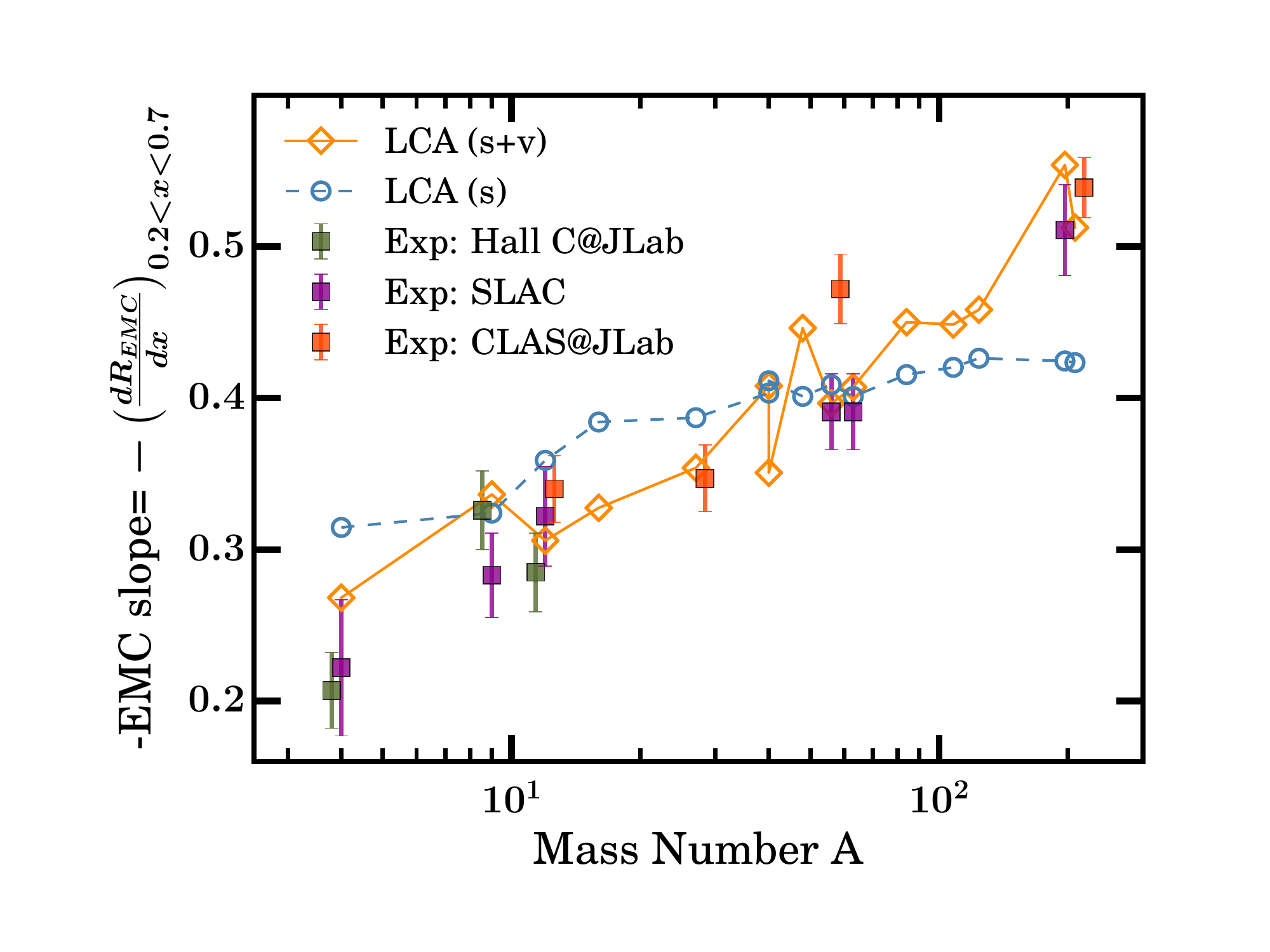}
\caption{LCA results for the size of the EMC effect based on a linear connection with the proton and neutron SRC scaling factors. The ``LCA (s)'' results (blue open circles) assume a generative mechanism that is isospin (flavor) blind ($m_1=0.103\pm0.002,m_2=0$ in Eq.~(\ref{eq:EMCwithsandv})). The ``LCA (s+v)''  results (orange open diamonds) assume  generative mechanisms that are the combination of an isospin-dependent and an isospin-independent component  ($m_1=0.0878\pm0.003, m_2=-0.229 \pm 0.030$ in Eq.~(\ref{eq:EMCwithsandv})). The  non-isoscalar corrected $b_2^{exp}(A)$ data for the EMC slopes are from the summarizing tables of Ref.~\cite{Schmookler:2019nvf}. Overlapping data points have been slightly displaced for the sake of clarity.} 
\label{fig:EMCresults}
\end{figure}

\subsection{Size of EMC effect}
\label{subsec:sizeEMCeffect}
We exploit the conjectured linear relationship between the $a_2$ coefficients (per nucleon modification relative to the deuteron) and the size of the EMC effect to connect the LCA predictions for $a_2^{p} (A)$ and $a_2^{n} (A)$ to the EMC data.  We suggest parametrizing the size of the EMC effect as defined in Eq.~(\ref{eq:sizeEMC}) in the following way
\begin{eqnarray} \label{eq:EMCwithsandv}
b_2^{exp}(A) & = &- \frac {d R_{EMC} (A,x)} {d x} 
\nonumber \\
& = & 
m_1 \left(
\frac{Z a_{2}^p(A) + N a_{2}^{n}(A)} {A} -1
\right) 
\nonumber \\ 
& & + m_2
\left(
\frac{Z a_{2}^p (A) - N a_{2}^{n}(A)} {A}
\right) \; ,
\end{eqnarray}
with $m_1$ and $m_2$ being parameters that are here determined from theory-experiment comparisons. By construction the deuteron has a vanishing EMC effect. 
In the above linear relationship that connects the measured size of the EMC effect $b_2^{exp}(A)$ to the computed SRC scaling factors for protons and neutrons, the first term (weight $m_1$) is reminiscent of an isospin blind generative mechanism. Indeed, in the first term of Eq.~(\ref{eq:EMCwithsandv}) the protons and neutrons contribute according to their weight $\frac{Z}{A}$ and $\frac{N}{A}$ in the total number of nucleons. For this reason, we refer to the term with weight $m_1$ as the ``isoscalar'' contribution. We refer to the second term (weight $m_2$) as the ``isovector'' contribution that could find its origin in an isospin-dependent generative mechanisms for the EMC effect. Within the framework of relativistic quark-level models of nuclear structure, it has been suggested that those so-called flavor-dependent or isovector nuclear effects influence the size of the EMC effect in nuclei with a neutron excess~\cite{Cloet:2009qs,Cloet:2015tha}. We remind that in the limit of vanishing proton-proton and neutron-neutron correlations in $N\ne Z$ nuclei, the $a_{2}^p (A)$ and $a_{2}^{n}(A)$ obey the relation of Eq.~(\ref{eq:limitpndominant}) and the $m_2$ term in the above equation vanishes. 
 
 In Fig.~\ref{fig:EMCresults} we compare the LCA predictions based on Eq.~(\ref{eq:EMCwithsandv}) with data for the EMC slopes without applying isoscalar corrections to those data. We use $\chi^2$ minimization to fit the computed $a_2^p(A)$ and $a_2^n(A)$ to 13 measured EMC slopes using the expression (\ref{eq:EMCwithsandv}).   We include the measured EMC slopes for the following nuclei (with $xm$ we denote that there are $x$ measurements for a particular target nucleus): $^4$He ($2m$), $^9$Be ($2m$), $^{12}$C ($3m$), $^{27}$Al ($1m$), $^{56}$Fe ($2m$), $^{63}$Cu ($1m$), $^{197}$Au ($1m$), $^{208}$Pb ($1m$). Accordingly, there are eight target nuclei included in the fit, including six asymmetric ones.  The best description of the data is obtained with the combination ($m_{1}=0.0878 \pm 0.003, m_2 = -0.229 \pm 0.030$). The values for the EMC slopes obtained with the Eq.~(\ref{eq:EMCwithsandv}) are contained in Table~\ref{tab:all_results}. The extracted value for $m_{1}=0.0878 \pm 0.003$ is in line with the quoted value $m_1 = 0.084 \pm 0.004$ in Ref.~\cite{Lynn:2019vwp} that is based on fit of $- \frac {d R_{EMC} (A,x)} {d x}$ versus $a_2^{exp}(A)$. The fit of~Ref.~\cite{Lynn:2019vwp} 
 includes all EMC data for light nuclei ($A\leq 12$) and the isoscalar-corrected EMC data for $^{56}$Fe and $^{197}$Au. The results presented in Ref.~\cite{Lynn:2019vwp} include solely the first term of the right-hand side of Eq.~(\ref{eq:EMCwithsandv}).   The EMC data for $A \le 63$ can be reasonably reproduced without inclusion of an isovector term ($m_1=0.103\pm0.002,m_2=0$), a result referred to as ``LCA(s)''.  Without inclusion of the isovector term, the EMC slope for $^4$He and $^{27}$Al tends to be over-predicted whereas  for the EMC slopes of $^{197}$Au and $^{208}$Pb the opposite is observed. Inclusion of the $m_2$ term results in a stronger variation of the EMC slopes across the nuclear mass table and results in an improved description of the data. Indeed, after including the $m_2$ term the reduced  $\chi^2$ is 2.66 whereas the fit with solely the $m_1$ term has a reduced $\chi^2$ of 7.3. Obviously, a more precise determination of the $m_2$ term requires more data and extended studies on asymmetric $\frac{N}{Z}>1$ nuclei. For example, the effect of including the $m_2$ term works very differently for the nuclei $^{40}$Ca and $^{48}$Ca. Power-counting arguments within the framework of effective-field theory ~\cite{Lynn:2019vwp} indicate that the isovector term $\frac{Z a_{2}^p (A) - N a_{2}^{n}(A)}{A}$ can be neglected relative to the isoscalar one $\frac{Z a_{2}^p (A) +  N a_{2}^{n}(A)}{A}$.  Inspecting Table~\ref{tab:all_results} we find that the isoscalar term is in the range 4.05-5.11, whereas the iosvector term is in the range $-0.84-0.00$. The isovector term is vanishing for $N=Z$ nuclei and reaches its largest values for the most asymmetric nuclei considered here: $-0.66$ for $^{208}$Pb and $-0.84$ for $^{197}$Au.  We stress that the above theory-experiment comparisons for the EMC slopes cannot shed light on underlying mechanisms that are due to non-SRC related medium modifications.  

%
%
\section{Conclusions}
\label{sec:conclusions}

The SRC scaling factors $a_2$ represent the relative probability of nucleon-pair SRC in a specific nucleus relative to the deuteron. They are conventionally expressed per nucleon and can be computed from the high-momentum properties of momentum probability distributions. 
In the framework of the low-order correlation operator approximation used throughout this work one can determine the pp, pn, nn, np SRC contributions to the  momentum probability distributions $P^A(p)$ for a specific nucleus A(N,Z). We have determined those contributions for a sample of 15 nuclei extending in mass number from He to Pb. Across that sample, that includes four symmetric and eleven asymmetric nuclei, there is relatively little variation in the computed $a_2$, with values in the range between 4.05 (for $^4$He) and $\approx 5.10$ (all studied nuclei with $A \ge 108$).  We find that the pn contribution to the SRC scaling factor is approximately 3 and that there are  non-negligible contributions from pp and nn correlations in the LCA approach. The pp and pn (nn and np) SRC contributions to the high-momentum probability distributions determine the proton (neutron) SRC scaling factors $a_2^p$ ($a_2^n$).    The $a_2^p$ and $a_2^n$ provide more detailed information on the abundance of nucleon-pair SRC than the ``total SRC scaling factor $a_2$''. For asymmetric $N>Z$ nuclei one systematically finds that  $a_2^p > a_2^n$ with deviations approaching 20\% for the most asymmetric nuclei in our study.   This means that in $N>Z$ nuclei the SRC induced medium modifications of the protons and the neutrons are substantially different. We have done robustness checks and used two different techniques to compute 
the SRC scaling factors $\left( a_2, a_2^p, a_2^n \right)$. For light and medium-heavy nuclei, the LCA predictions for the SRC scaling factors are in line with those from ab-initio approaches and the values extracted from inclusive electron scattering under selected conditions.      

In the LCA framework we can shed light on the validity of the $A$-to-$d$ factorization of the momentum probability distributions by studying the ratios  $\frac{P^{A}_{NN^{\prime}}(p)}{P^{d}(p)}$.  For $NN^{\prime}$=pn and np the $A$-to-$d$ factorization is very well realized at $p \gtrsim 3.5$~fm$^{-1}$. For the pp and nn correlations, on the other hand, the $A$-to-$d$ factorization of $\frac{P^{A}_{NN^{\prime}}(p)}{P^{d}(p)}$ is only approximate but indications for a plateau are visible for $p \gtrsim 4.0$~fm$^{-1}$. We have expressed the measured size of the EMC effect in terms of the computed proton and neutron SRC scaling factors. The measured size of the EMC effect displays a stronger variation across the nuclear mass table than the SRC scaling factor and larger EMC effects are observed in nuclei with a neutron excess. These qualitative features can be captured in terms of a linear relationship between the size of the EMC effect and the computed proton and neutron SRC scaling factors that includes both an isoscalar and an isovector term.

\section*{ACKNOWLEDGMENTS}
The computational resources (Stevin Supercomputer Infrastructure) and services used in this work were provided by the VSC (Flemish Supercomputer Center), funded by Ghent University, FWO and the Flemish Government – Department EWI. We thank O.~Hen for many insightful discussions.

 %

\end{document}